\documentclass[useAMS]{mn2e}
\usepackage{psfig,epsfig}
\usepackage{graphicx}
\usepackage{amssymb}
\usepackage{epstopdf}

\begin{document}
\def\simlt{\mathrel{\rlap{\lower 3pt\hbox{$\sim$}}
        \raise 2.0pt\hbox{$<$}}}
\def\simgt{\mathrel{\rlap{\lower 3pt\hbox{$\sim$}}
        \raise 2.0pt\hbox{$>$}}}
	
\title[FIR properties of radio-selected sources]{A census of radio-selected AGN on the COSMOS field and of their FIR properties}

\author[Manuela Magliocchetti et al.]
{\parbox[t]\textwidth{M. Magliocchetti$^{1}$, P. Popesso$^{2}$, M. Brusa$^{3,4}$, M. Salvato$^{5}$\\
} \\
{\tt $^1$ INAF-IAPS, Via Fosso del Cavaliere 100, 00133 Roma,
  Italy}\\
 {\tt $^2$ Excellence Cluster, Boltzmannstr. 2, D85748, Garching, Germany}\\
  {\tt $^3$ Dipartimento di Fisica e Astronomia, Universita' di Bologna, Via Gobetti 93/2, 40129, Bologna, Italy}\\
    {\tt $^4$ INAF-Osservatorio Astronomico di Bologna, Via Gobetti 93/3, 40129, Bologna, Italy}\\
     {\tt $^5$ Max Planck Institut f\"ur extraterrestrische Physik (MPE),
  Postfach 1312,  D85741, Garching, Germany}\\  
   }
 \maketitle	

\begin{abstract}
We use the new catalogue by Laigle et al. (2016) to provide a full census of VLA-COSMOS radio sources.  We identify 90\% of such sources and sub-divide them into AGN and star-forming galaxies on the basis of their radio luminosity. The AGN sample is {\it complete} with respect to radio selection at all $z\simlt 3.5$. Out of 704 AGN, 272 have a counterpart in the {\it Herschel} maps. By exploiting the better statistics of the new sample, we confirm the results of Magliocchetti et al. (2014): the probability for a radio-selected AGN to be detected at FIR wavelengths is both a function of radio luminosity and redshift, whereby powerful sources are more likely FIR emitters at earlier epochs. Such an emission is due to star-forming processes within the host galaxy. FIR emitters and non-FIR emitters only differentiate in the $z\simlt 1$ universe. At higher redshifts they are indistinguishable from each other, as there is no difference between FIR-emitting AGN and star-forming galaxies. 
Lastly, we focus on radio AGN which show AGN emission at other wavelengths. We find that MIR  emission is mainly associated with ongoing star-formation and with sources which are smaller, younger and more radio luminous than the average parent population. 
X-ray emitters instead preferentially appear in more massive and older galaxies. We can therefore envisage an evolutionary track whereby the first phase of a radio-active AGN and of its host galaxy is associated with MIR emission, while at later stages the source becomes only active at radio wavelengths and possibly also in the X-ray.
\end{abstract}

\section{Introduction}

While in the recent years there has been a growing interest of the scientific community in extra-galactic radio objects, and very ambitious programs like the planned Square Kilometer Array (SKA, Carilli et al. 2004)  or its precursors ASKAP (Australian SKA pathfinder, Johnston S. et al. 2007) and MeerKAT (Jonas J.L. 2009) will soon see their first light, investigating the very nature of these sources is not an easy task. They are indeed a mixed bag of astrophysical objects: from the powerful radio-loud QSOs and FRII (Fanaroff \& Riley 1974) galaxies to the weaker FRI,  low-excitation galaxies and  star-forming galaxies, whose contribution to the total radio counts becomes predominant at the sub-mJy level (see e.g. Magliocchetti et al. 2000; Prandoni et al. 2001; Bonzini et al. 2013, Bonzini et al. 2015; Padovani et al. 2015). Discerning amongst them is only possibile via photometric and, whenever possibile, spectroscopic follow-ups and to this aim a series of deep-field radio surveys have been performed on very well studied cosmological fields such as COSMOS (Schinnerer et al. 2004; 2007; 2010), GOODS-North (Morrison et al. 2010), VIDEO-XMM3 (McAlpine et al. 2013), VVDS (Bondi et al. 2003), Subaru/XMM (Simpson et al. 2013) and the Extended Chandra Deep Field (Mao et al. 2011 and Bonzini et al. 2012).

Following the launch of the {\it Spitzer} and {\it Herschel} satellites, radio sources have also been investigated at Mid-Infrared (MIR) (e.g. Appleton et al. 2004; Boyle et al. 2007; Magliocchetti, Andreani \& Zwaan 2008; Garn et al. 2009;  Leipski et al. 2009; De Breuck et al. 2010; Norris et al. 2011 amongst the many) and Far-Infrared (FIR) wavelengths (e.g. Seymour et al. 2011; Del Moro et al. 2013; Magliocchetti et al. 2014, 2016). 
These studies allowed to probe the sub-population of star-forming galaxies up to $z\sim 3$, and also provided invaluable information on the central engine responsible for radio (and MIR) AGN emission.

In more detail, Magliocchetti et al. (2014) for the first time investigated the FIR properties of AGN selected at 1.4 GHz of {\it all} radio luminosities (except for the "monsters" which do not appear in small and deep fields) and at {\it all} redshifts $z\simlt 3.5$. 
This was done by using the very deep catalogue obtained at 1.4 GHz on the COSMOS field by Schinnerer et al. (2004; 2007; 2010) and Bondi et al. (2008); AGN were selected solely on the basis of their radio-luminosity and FIR information was provided by the PEP survey (Lutz et al. 2011, 2014). Magliocchetti et al. (2016) then extended this previous analysis to deeper FIR fields such as GOODS-North, GOODS-South and the Lockman Hole. 

The present work is the third one of the above series and re-analyses the radio and FIR properties of radio-selected sources in the COSMOS field by making use of the very recently published catalogue of redshifts by Laigle et al. (2016). 
In fact, the COSMOS field (Scoville et al. 2007) covers a large ($\sim 2$ deg$^2$) area and it is observed with very deep (AB$=25-26$) multi-wavelength data, including imaging in 18 intermediate band filters from Subaru (Taniguchi et al. 2007), which allow to pinpoint emission/absorption lines in the SEDs, and NIR/MIR data from UltraVISTA (McCracken et al. 2012) and IRAC (SPLASH Survey; Capak et al. in prep). The photometry is homogenized and blending is also taken into account, making the quality of the data, the photometric redshifts and the stellar masses in the Laigle et al. (2016) catalogue among the best available. The availability of reliable photometric redshifts is not limited to normal galaxies but it also assured for the X-ray sources detected by Chandra in a deep and homogeneous manner (Civano et al. 2016, Marchesi et al. 2016).

Indeed, while the work by Magliocchetti et al. (2014) was based on the catalogue by Ilbert et al. (2013), which only provided redshifts for about 65\% of the radio sources from  the VLA-COSMOS survey, the Laigle et al. (2016) dataset provides information for more than 90\% of such objects. This is due to the fact that, unlike the Capak et al. (2007) and Ilbert et al. (2010) works which relied on i-band selection, the Laigle et al. (2016) catalogue is based on a near-infrared $zYJHK$ selection which allows to sample much redder objects and therefore probe higher redshift sources. This extremely high completeness then allows to reinvestigate the radio and FIR properties of COSMOS radio-selected AGN and draw conclusions at much higher confidence levels. 

Furthermore, galaxies in the Laigle et al. (2016) catalogue are also provided with a flag which indicates whether the source shows signature for AGN emission in a number of wavebands. This information will then be used throughout the present work to assess whether there are systematic differences amongst radio-active AGN which also emit in the MIR or X-ray bands and to envisage an evolutionary connection between these different classes of sources.

Throughout the work we will assume a $\Lambda$CDM cosmology with $H_0=70 \: \rm km\:s^{-1}\: Mpc^{-1}$ ($h=0.7$), $\Omega_0=0.3$,  $\Omega_\Lambda=0.7$ and $\sigma^m_8=0.8$. Masses for the sources under examination come from the Laigle et al. (2016) catalogue and are calculated using the Bruzual \& Charlot (2003) templates and adopting  a Salpeter (1955) Initial Mass Function  (IMF). 

\begin{figure}
\includegraphics[scale=0.4]{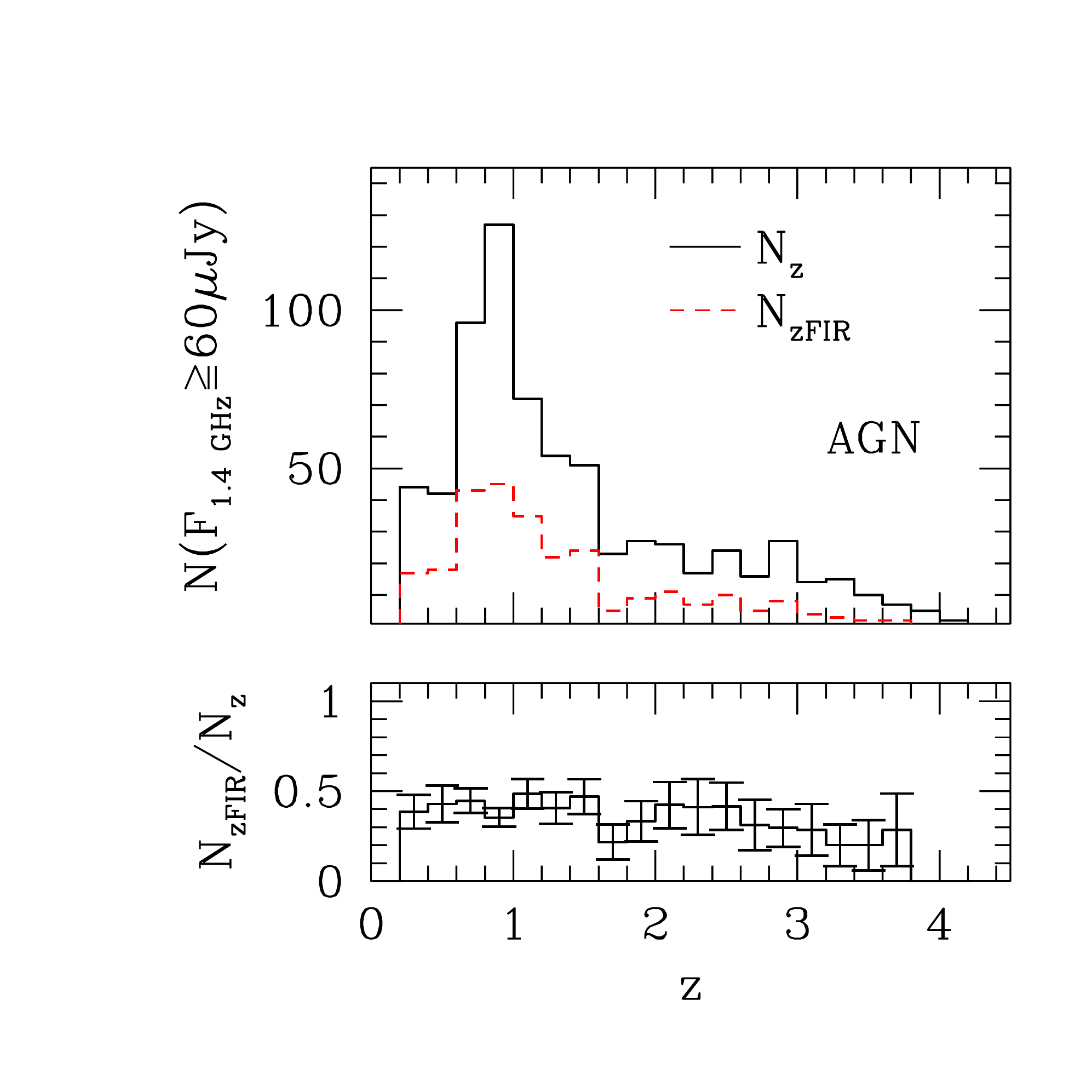}
\caption{Redshift distribution of COSMOS-VLA radio-selected AGN with fluxes F$_{\rm 1.4 GHz}\ge 0.06$ mJy. The AGN sample is complete up to $z\sim 3.5$. The solid line represents all sources irrespective of their FIR emission, while the dashed line indicates those objects which are also FIR emitters.  The bottom panel highlights the ratio between the two quantities. Errorbars correspond to 1$\sigma$ Poissonian estimates.
\label{fig:idsvsz}}
\end{figure}

\section{The Radio-Infrared Master Catalogue}

The main properties of the radio and FIR catalogues are extensively described in Magliocchetti et al. (2014). Here we report a brief summary. The VLA-COSMOS Large Project observed the 2 deg$^2$ of the  COSMOS field at 1.4 GHz  (Schinnerer et al. 2004; Schinnerer et al. 2007). The catalogue adopted in our work is that presented in Bondi et al. (2008), which comprises 2382 sources selected above a 1.4 GHz integrated flux of 60 $\mu$Jy. 

The COSMOS region has been observed down to $\sim 4$ mJy at 100$\mu$m and $\sim 7$ mJy at 160$\mu$m by the PACS (Poglitsch et al. 2010) instrument onboard the {\it Herschel} Space Observatory (Pilbratt et al. 2010) as a part of the PACS Evolutionary Probe (PEP,  D. Lutz et al. 2011) Survey.
Infrared counterparts to F$_{1.4 \rm GHz}\ge 60\mu$Jy VLA-COSMOS sources have been found by a simple matching technique between the radio and the COSMOS-PEP catalogues. 
By adopting the same criteria of Magliocchetti et al. (2014), we chose as maximum separation values 4$^{\prime\prime}$ at 100$\mu$m and 5$^{\prime\prime}$ at 160$\mu$m. 
We find that 1063 VLA-COSMOS sources (corresponding to 44\% of the parent sample) have a counterpart at 100$\mu$m and 1100 (corresponding to 46\%) have a counterpart at 160$\mu$m. 
The total number of F$_{1.4 \rm GHz}\ge 60 \mu$Jy VLA-COSMOS sources with an infrared counterpart either at 100$\mu$m or at 160$\mu$m is 1219, corresponding to 51\% of the original sample. 

Finally, in order to provide the overwhelming majority of radio sources with a redshift determination, in this work we cross correlated the above sample with the Laigle et al. (2016)  catalogue which provides reliable photometric redshifts ($\sigma_{NMAD} = 0.01$ for galaxies brighter than $I=22.5$) for COSMOS galaxies, with only a handful of outliers. When possible, we used spectroscopic redshifts available within the COSMOS collaboration.

Given the high positional accuracy of both the radio and the optical-near infrared surveys, we fix the matching radius to 1 arcsec. 
This procedure provides redshift estimates for 2123 radio sources (out of which 1199 are spectroscopic), corresponding to $\sim 90$ per cent of the parent sample. This has to be compared with our previous work where only $\sim 65$  per cent of the VLA-COSMOS objects were endowed with a redshift determination. 
Of the 2123 radio sources with redshift of the present work, 1173 also have a counterpart on the {\it Herschel} maps. This corresponds to 96\% of FIR-detected galaxies.

\section{AGN selection via radio luminosity}

\begin{figure*}
\includegraphics[scale=0.4]{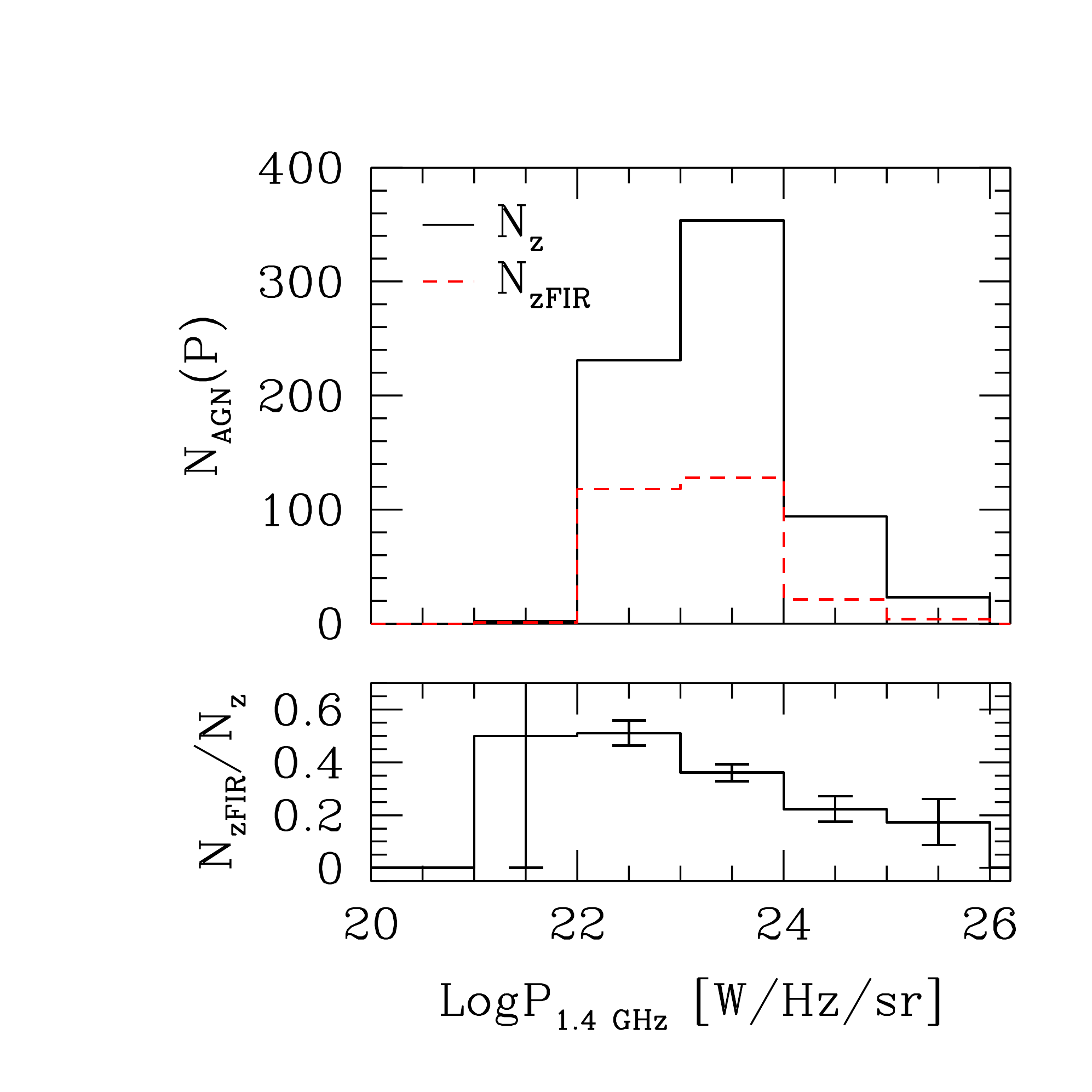}
\includegraphics[scale=0.4]{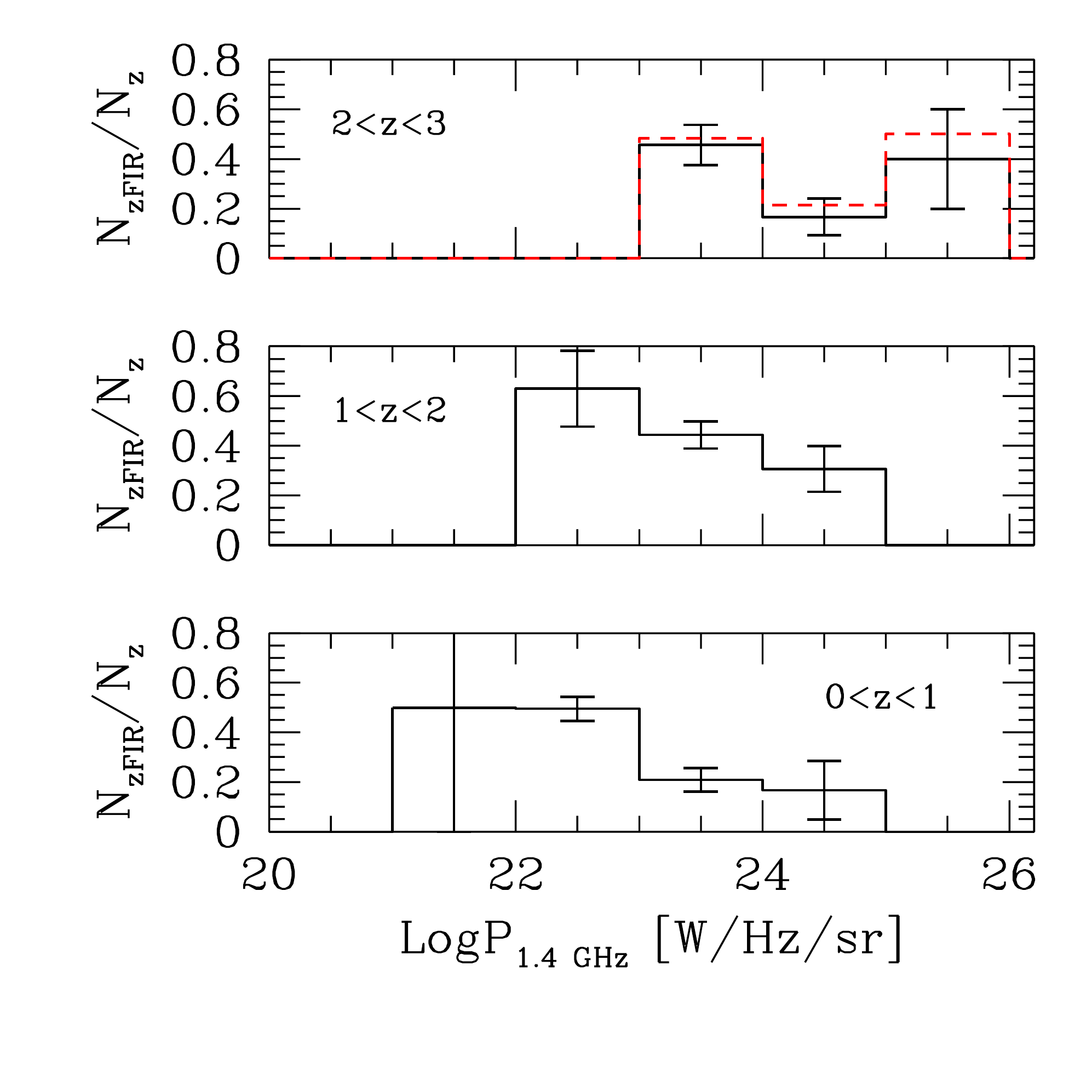}
\caption{Left-hand panel: Distribution of radio powers for F$_{\rm 1.4 GHz}\ge 0.06$ mJy COSMOS-VLA radio-selected AGN. The solid line represents all sources, irrespective of their FIR emission, while the dashed line indicates those objects which are also FIR emitters. The bottom histogram highlights the ratio between the two quantities.  Right-hand panel: distribution of the fraction of radio-selected AGN in the COSMOS-VLA survey which are also FIR emitters as a function of radio luminosity as seen in three different redshift intervals. 
The dashed histogram in the top panel represents the results obtained if we restrict to sources in the redshift range $z=[2-2.5]$. In both panels, errorbars correspond to $1\sigma$ Poissonian estimates. 
\label{fig:idsvsP}}
\end{figure*}

\begin{figure*}
\includegraphics[scale=0.4]{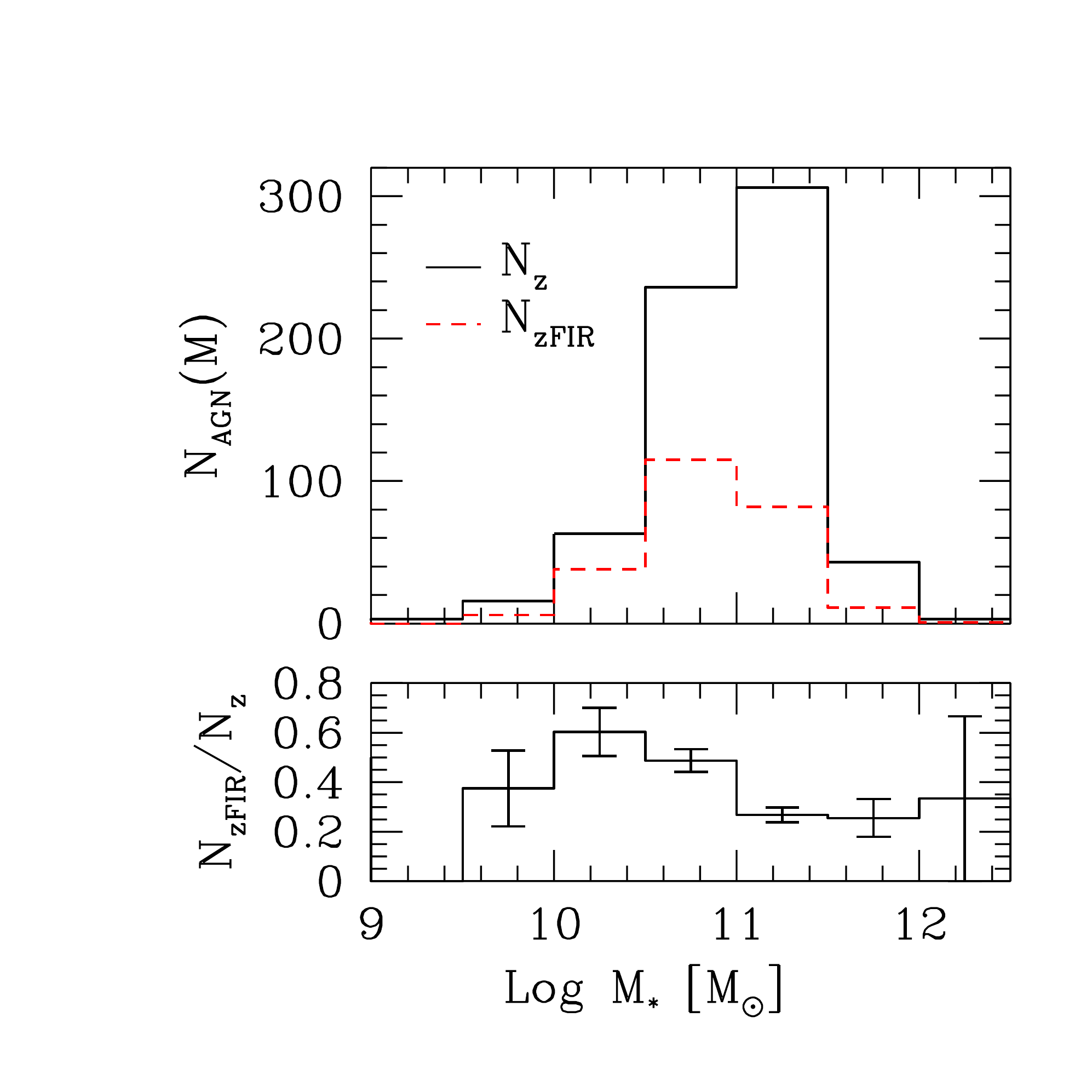}
\includegraphics[scale=0.4]{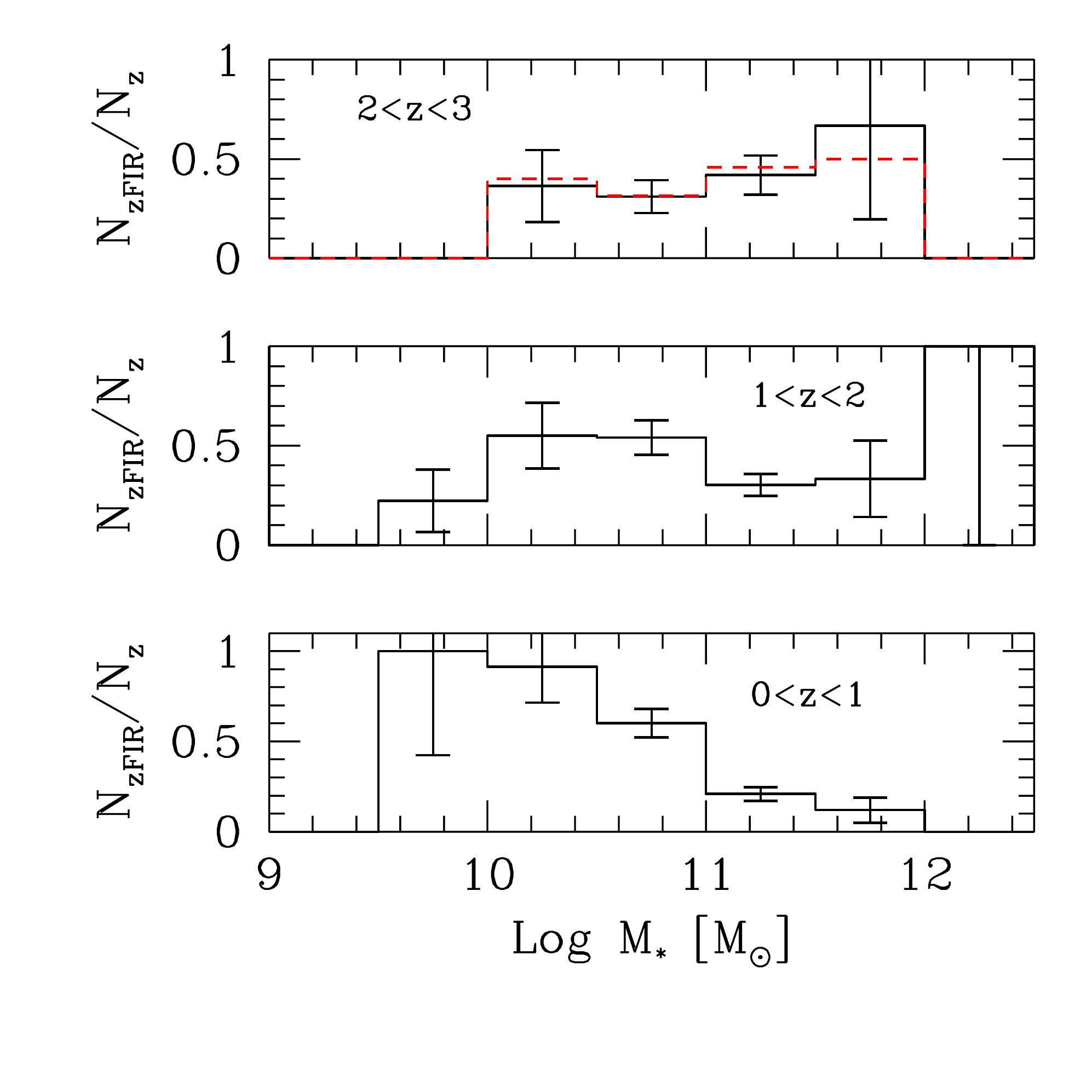}
\caption{Left-hand panel: Distribution of stellar masses for  COSMOS-VLA radio-selected AGN brighter than F$_{\rm 1.4 GHz}\ge 0.06$ mJy. The solid line represents all sources irrespective of their FIR emission, while the dashed line indicates those objects which are also FIR emitters. The bottom histogram highlights the ratio between these two quantities. Right-hand panel: Distribution of the fraction of radio-selected AGN in the COSMOS-VLA survey which are also FIR emitters as a function of stellar mass as seen in three different redshift intervals. The dashed histogram in the top panel represents the results obtained if we restrict to sources in the redshift range $z=[2-2.5]$. In both panels, errorbars correspond to $1 \sigma$ Poissonian estimates.
\label{fig:idsvsmass}}
\end{figure*}

\begin{figure*}
\includegraphics[scale=0.4]{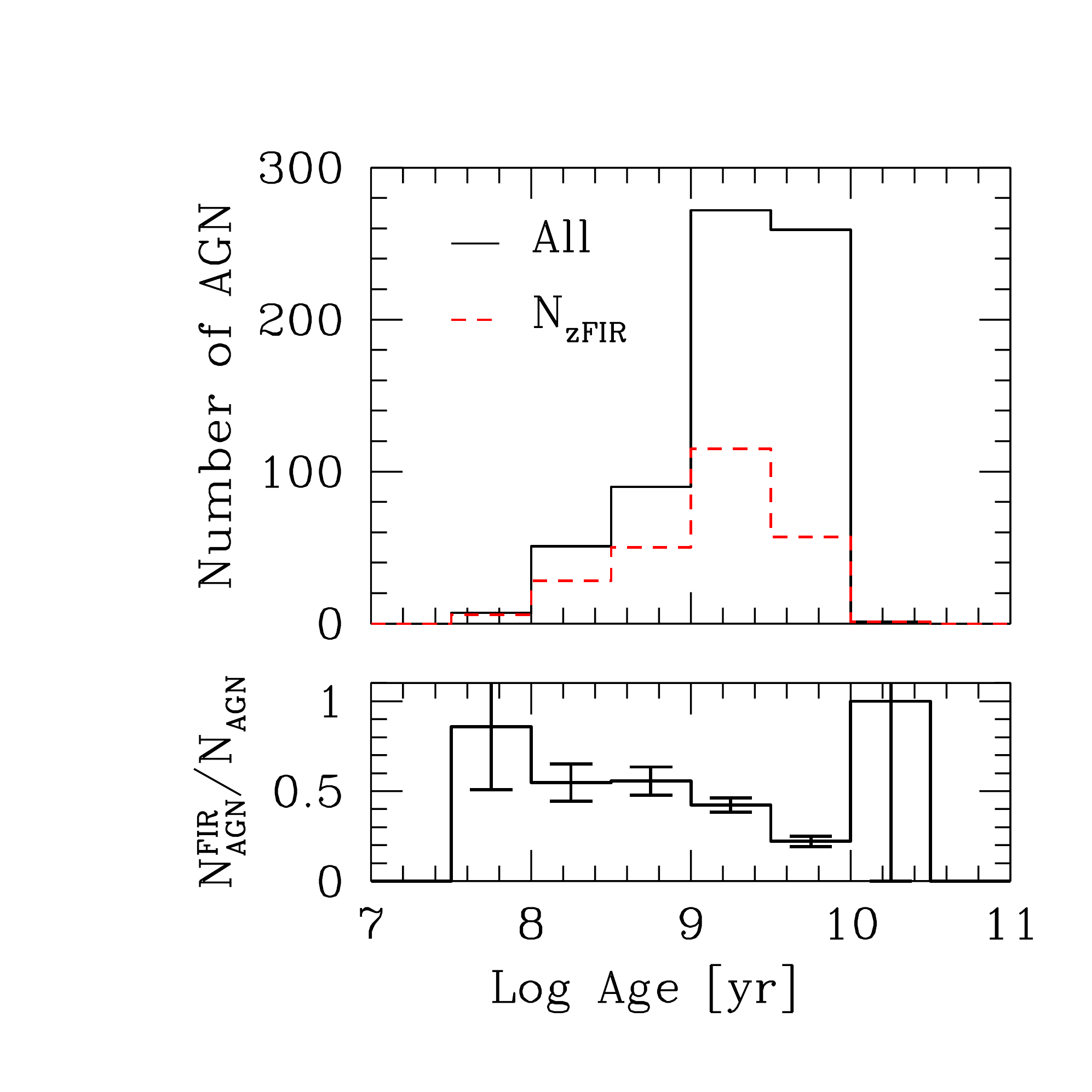}
\includegraphics[scale=0.4]{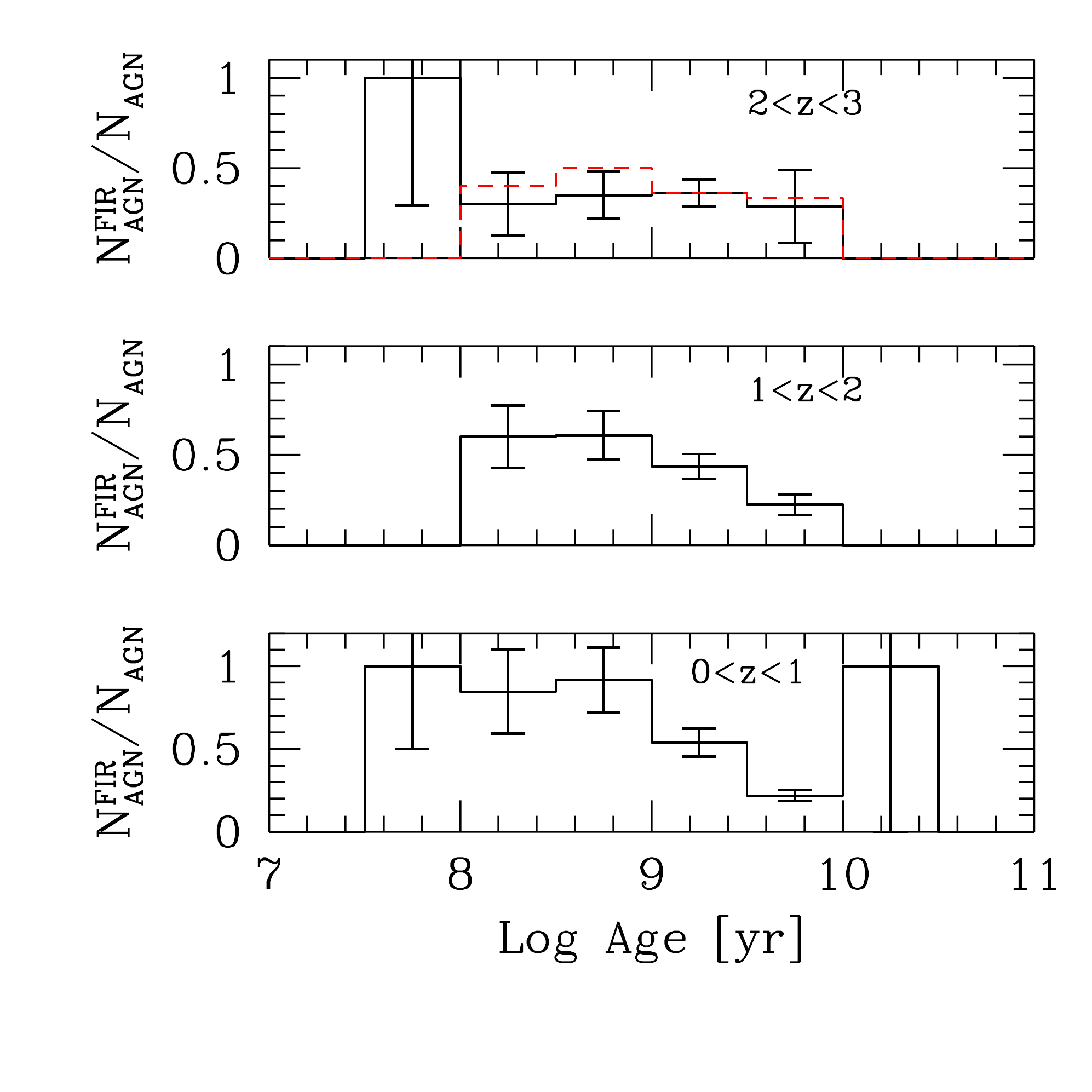}
\caption{Left-hand panel: Distribution of ages for  the hosts of COSMOS-VLA radio-selected AGN brighter than F$_{\rm 1.4 GHz}\ge 0.06$ mJy. The solid line represents all sources irrespective of their FIR emission, while the dashed line indicates those objects which are also FIR emitters. The bottom histogram highlights the ratio between these two quantities. Right-hand panel: Distribution of the fraction of radio-selected AGN in the COSMOS-VLA survey which are also FIR emitters as a function of galaxy age as seen in three different redshift intervals. The dashed histogram in the top panel represents the results obtained if we restrict to sources in the redshift range $z=[2-2.5]$. In both panels, errorbars correspond to $1 \sigma$ Poissonian estimates.
\label{fig:tau}}
\end{figure*}

A tricky point in the process of identifying extragalactic sources observed in monochromatic radio surveys is that of distinguishing between radio emission of AGN origin and that instead due to star forming processes.
The approach we adopt here is that introduced by Magliocchetti et al. (2014) and already used in Magliocchetti et al. (2016). This is based on the  results of McAlpine, Jarvis \& Bonfield (2013) who provide luminosity functions at 1.4 GHz for the two classes of AGN and star-forming galaxies.  Their results show that the radio luminosity P$_{\rm cross}$ beyond which  AGN-powered galaxies become the dominant radio population scales with redshift roughly as
\begin{eqnarray}
\rm Log_{10}P_{\rm cross}(z)=\rm Log_{10}P_{0,\rm cross}+z,
\label{eq:P}
\end{eqnarray}
at least up to $z\sim 1.8$. $P_{0,\rm cross}=10^{21.7}$[W Hz$^{-1}$ sr$^{-1}$] is the value which is found in the local universe and which roughly coincides with the break in the radio luminosity function of star-forming galaxies (cfr Magliocchetti et al. 2002; Mauch \& Sadler 2007). Beyond this value,  their  luminosity function steeply declines, and the contribution of star-forming galaxies to the total radio population is drastically reduced to a negligible percentage (Magliocchetti et al. 2002; Mauch \& Sadler 2007). 

We then distinguished between AGN-powered galaxies and star-forming galaxies by means of equation (\ref{eq:P}) for $z\le 1.8$ and by fixing $\rm Log_{10}P_{\rm cross}(z)=23.5$ [W Hz$^{-1}$ sr$^{-1}$] at higher redshifts. 
This procedure identifies 704 AGN (corresponding to 33\% of the total radio population) and 1419 star-forming galaxies. With respect to our earlier work, these numbers imply an increase of a third of sources in both the AGN and the SF samples and allow to draw conclusions on the properties of radio-selected sources with and without a FIR counterpart on much more solid grounds. Also, note that, due to the adopted selection criteria and thanks to the depth of the VLA-COSMOS survey,  the AGN samples is {\it complete} with respect to radio selection at all redshifts $z\simlt 3.5$, i.e. the considered sample includes  {\it all} radio-emitting AGN selected at 1.4 GHz in the COSMOS field and endowed with a redshift determination $z\simlt 3.5$.  272 sources classified as AGN and 901 sources classified as star-forming galaxies also show up in the {\it Herschel} maps.
 
The redshift distribution of radio-selected AGN is presented in Figure \ref{fig:idsvsz}. The solid histogram shows the distribution of all AGN, independent of their FIR emission, while the dashed (red) histogram represents that of those AGN which also appear in the {\it Herschel} maps. The bottom panel highlights the ratio between these two quantities. Errorbars correspond to 1$\sigma$ Poissonian estimates. As already seen in Magliocchetti et al. (2014), also in the present case we have that the redshift distribution of radio-selected AGN presents a marked peak at a redshift $z\simeq 1$. However, the better statistics of the present data allow to identify a prominent tail which extends up to $z\simeq 3.5-4$ and which in Magliocchetti et al. (2014) only showed up as a secondary peak centred around $z\sim 2.5$. There is no functional difference between the distribution of the parent AGN population and that which corresponds to AGN with a FIR counterpart. One is the scaled version of the other, and the ratio between these two quantities is roughly constant and equal to the value of $\sim 0.5$ throughout the whole redshift range probed by our data. 

Some more information on the sources under exam can be provided by investigations of the distribution of their radio luminosities, both in the presence and in absence of FIR emission. This is shown in Figure \ref{fig:idsvsP}, whereby the left-hand plot shows the distribution of sources at all redshifts, while the one on the right-hand side that of sources divided into redshift intervals. The left-hand plot of Figure  \ref{fig:idsvsP} clearly shows that the global distribution of radio-selected AGN (solid, black line) has a marked peak in the radio luminosity interval $\rm Log_{10}P_{\rm cross}(z)=23-24$ [W Hz$^{-1}$ sr$^{-1}$]. Beyond that value, there is a sharp drop in the number of AGN of higher luminosities. A very similar drop is also observed in the distribution of radio-selected AGN which are also associated with FIR emission (red, dashed line). However, the distribution of these sources at lower radio powers does not feature the same sharp peak in the range $\rm Log_{10}P_{\rm cross}(z)=[22-24]$ [W Hz$^{-1}$ sr$^{-1}$]. The net result of these two trends is that the fraction of radio-selected AGN which also emit at FIR wavelengths (shown in the bottom panel of the left-hand plot) is a strong function of radio luminosity. There is a clear trend which indicates that the number of FIR emitters monotonically decreases with increasing radio luminosity. This decrement is rather significant as the fraction of FIR emitters goes from $\sim 60$\% of the total radio-AGN population at luminosities $\rm Log_{10}P_{\rm cross}(z)\le 23$ [W Hz$^{-1}$ sr$^{-1}$] down to $\sim 20$\% for luminosities $\rm Log_{10}P_{\rm cross}(z)\ge 24$ [W Hz$^{-1}$ sr$^{-1}$]. However, interestingly enough, the right-hand plot of Figure \ref{fig:idsvsP} clearly shows that this marked decrement is only true in the relatively low ($z\simlt 2$) redshift universe, and gradually loses its importance when we move from local to more distant sources. Indeed, at higher redshifts we find that the fraction of FIR emitters is independent of radio luminosity, i.e. that all high-redshift radio-selected AGN have the same chances of being associated with FIR emission, independent of their radio luminosity. This result confirms those of Magliocchetti et al. (2016), while it partially contradicts that of Magliocchetti et al. (2014) who still found a dependence on radio luminosity also at the highest redshifts, as this was masked by relatively large uncertainties. In this respect, it is worth mentioning that the value of $\sim 0.4-0.5$ found in this work for the fraction of FIR emitters at $z\simgt 2$ is determined by the relatively shallow {\it Herschel} observations on the COSMOS field. Indeed, in the case of deeper observations like those on the two GOODS fields and on the Lockman Hole, one finds that such a percentage reaches the value of $100$\% (Magliocchetti et al. 2016).

\begin{figure}
\includegraphics[scale=0.4]{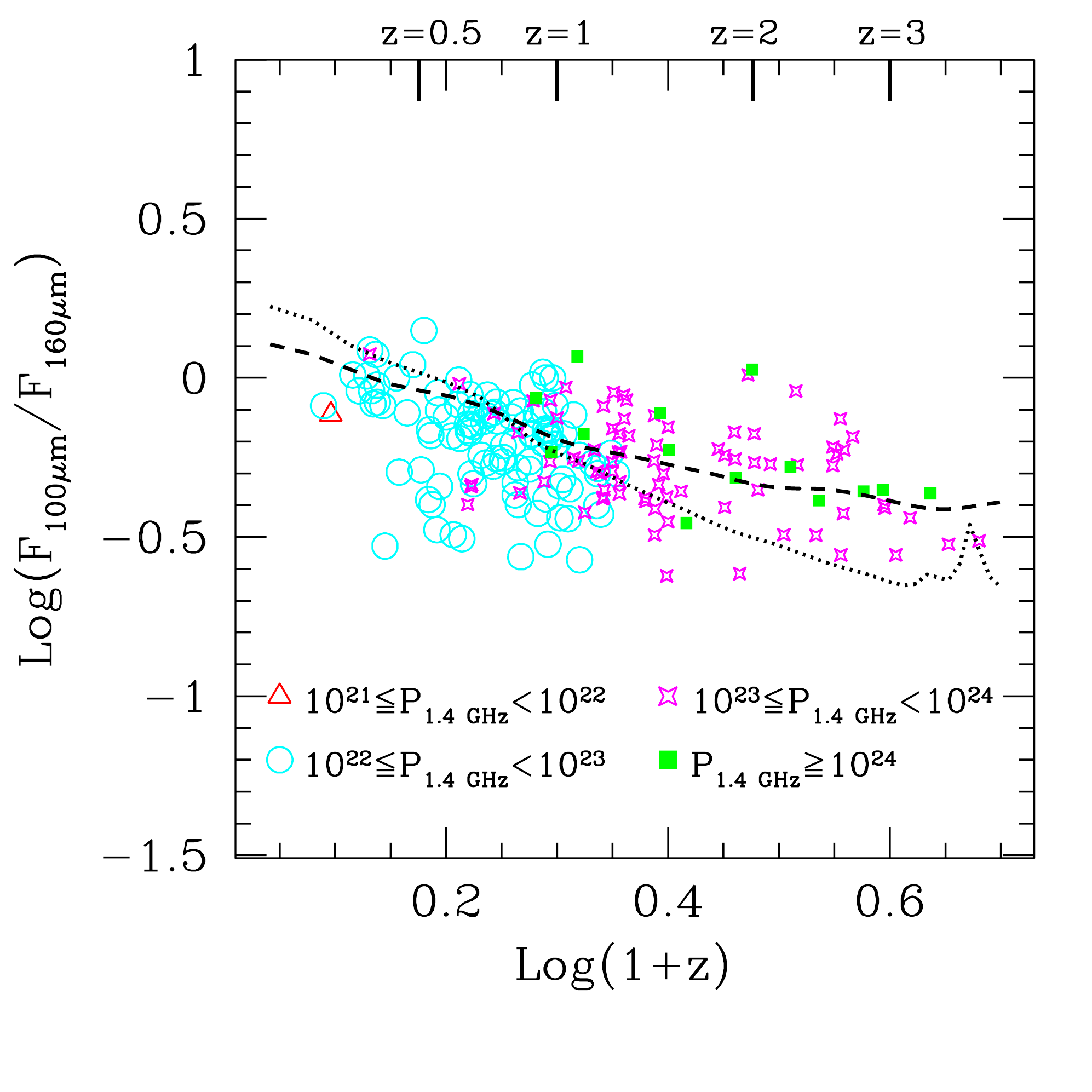}
\caption{Ratio between F$_{100 \rm \mu m}$ and F$_{160 \rm \mu m}$ fluxes as a function of redshift for radio-selected AGN with a FIR counterpart from the PEP survey. Different symbols correspond to different intervals for the radio luminosity (measured in [W Hz$^{-1}$ sr$^{-1}$]).The dashed line represents the trend obtained for the SED of M82, while the dotted one corresponds to Arp220 (see text for details).
\label{fig:q100_160}}
\end{figure}	

Another interesting feature that can be appreciated in the right-hand plot of Figure \ref{fig:idsvsP} is that the chances for an AGN of fixed radio luminosity to be a FIR emitter sensibly increase when moving from the lowest to the highest redshifts probed by our analysis. So, for instance, a source with P$_{1.4 \rm GHz}\simeq 10^{23}$ [W Hz$^{-1}$ sr$^{-1}$] will only have 
a $\sim 20$\% probability of being a FIR emitter at $z\le1$, while this percentage rises to $\sim 40$\% in the redshift range $z=[1-2]$, up to $\sim 50$\% for $z=[2-3]$ and a source with P$_{1.4 \rm GHz}\simeq 10^{25}$ [W Hz$^{-1}$ sr$^{-1}$] will have 
a $0$\% probability  of being a FIR emitter both at  $z\le1$ (zero objects out of one) and $1\le z\le 2$ (zero objects out of 11), while the percentage rises to $\sim 40$\% (four objects out of ten) at  $z=[2-3]$. \\
This trend, already highlighted by the work of Magliocchetti et al. (2014) and confirmed here on the more solid grounds provided by an almost complete optical identification of all radio-selected AGN, implies that   {\it the probability for a radio-selected AGN to be active in the FIR is both a function of radio power and redshift, whereby powerful sources are more likely to emit at FIR wavelengths at higher redshifts}.

\begin{figure*}
\includegraphics[scale=0.4]{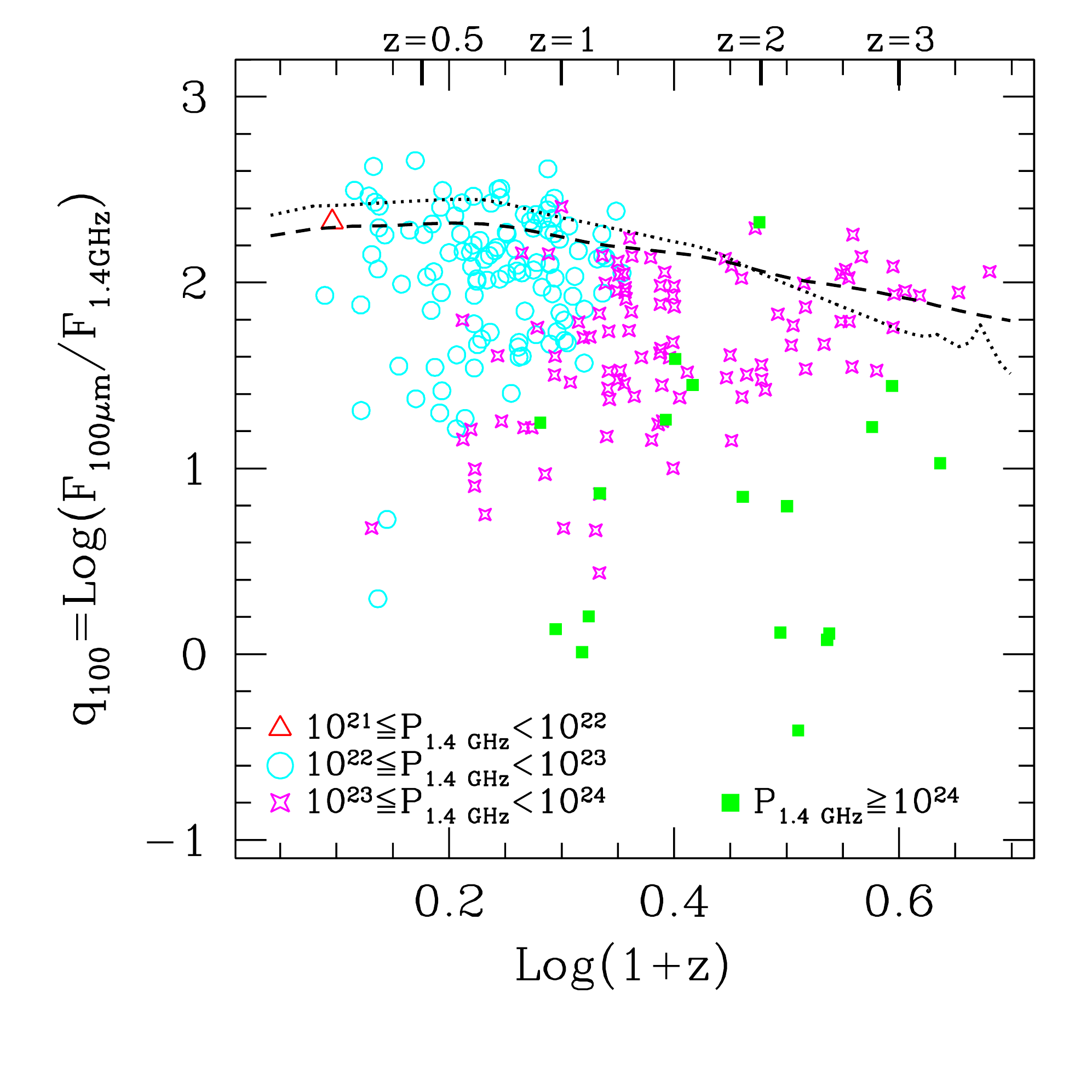}
\includegraphics[scale=0.4]{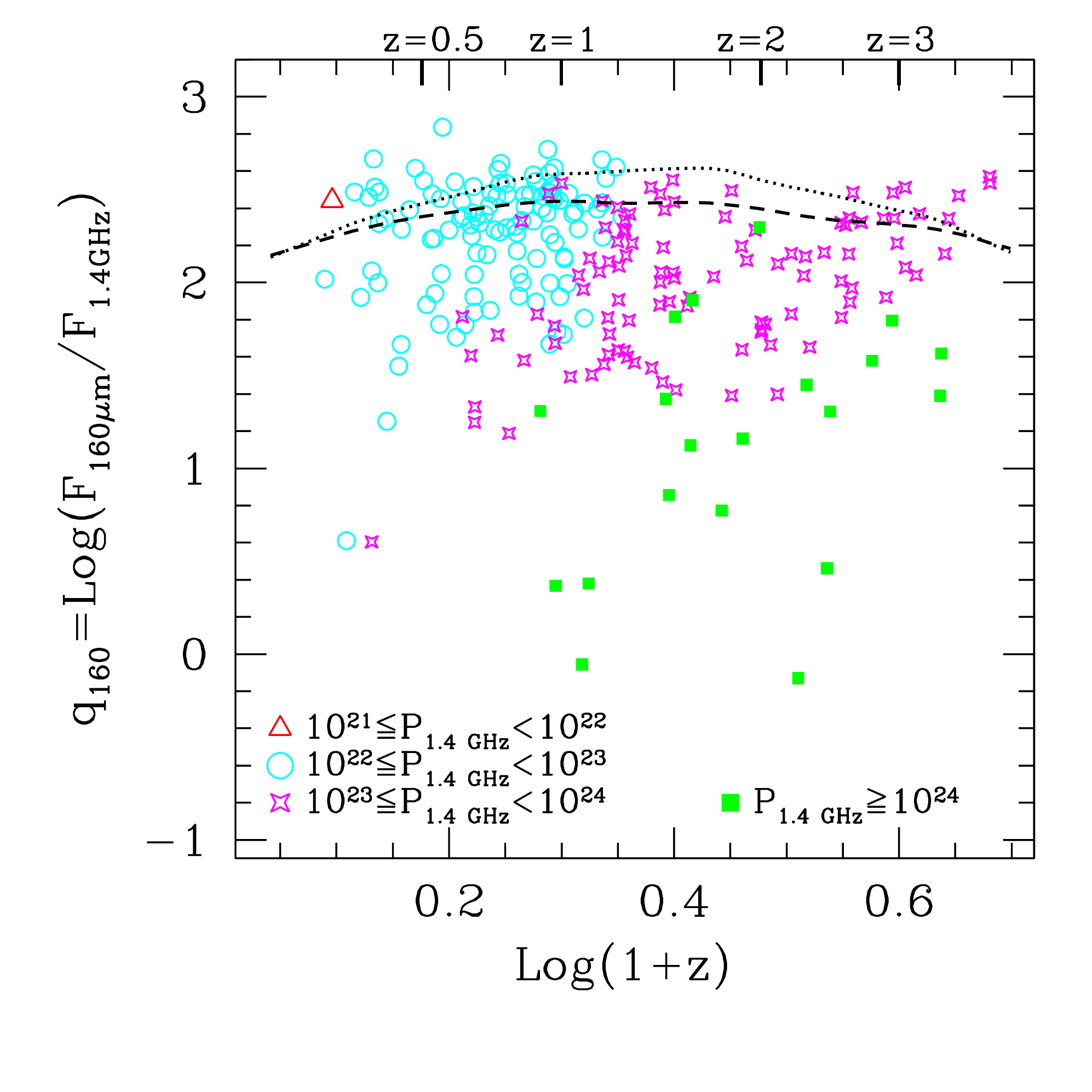}
\caption{Ratio between F$_{100 \rm \mu m}$ and F$_{1.4 \rm GHz}$ fluxes (left-hand panel) and F$_{160 \rm \mu m}$ and F$_{1.4 \rm GHz}$ fluxes (right-hand panel) as a function of redshift for radio-selected AGN with a FIR counterpart from the PEP survey. Different symbols correspond to different intervals for the radio luminosity (measured in [W Hz$^{-1}$ sr$^{-1}$]).The dashed lines represent the trends obtained for the SED of M82, while the dotted ones correspond to Arp220 (see text for details).
\label{fig:q100}}
\end{figure*}

Most of the above radio-selected AGN (696 out of 704) are also endowed with a mass estimate from the Laigle et al. (2016) catalogue\footnote{Note that the masses computed in Laigle et al. (2016) do not include an AGN component in the SED-fitting procedure and are calculated at the photometric redshift values. Both factors could in principle affect their estimates. Further work needs to be developed in order to address this point more properly.}. Figure \ref{fig:idsvsmass} presents the distribution of such sources as a function of stellar mass. As it was in the previous case, the left-hand plot presents the distribution of masses at all redshifts, while that on the right-hand side illustrates the trends in different redshift intervals. 
Radio-selected AGN are on average associated with galaxies with a large stellar mass content. From the left-hand plot of Figure  \ref{fig:idsvsmass} it is clear that more than 90\% of the sources has stellar masses M$_*\ge 10^{10}$ M$_\odot$ and more than 50\% of them is further associated with galaxies of masses M$_*\ge 10^{11}$ M$_\odot$. Those radio-selected AGN which are also FIR emitters instead present on average slightly lower values for the masses: their distribution (highlighted by the red, dashed line in Figure \ref{fig:idsvsmass}) peaks between $10^{10.5}$ M$_\odot$ and $10^{11}$ M$_\odot$, rather than in the range $[10^{11}-10^{11.5}]$ M$_\odot$ observed in the distribution of the whole radio AGN population. The net effect of these two different behaviours is that the fraction of radio-selected AGN associated with FIR emission (represented in the bottom panel of the plot on the left-hand side of Figure \ref{fig:idsvsmass}) remains roughly constant and equal to the value of $50-60$\% in the relatively low-mass, M$_*\simlt 10^{11}$ M$_\odot$, regime, while it drastically drops to the value of about 20-30\% for masses higher than the previous value. This implies that FIR emission in radio-selected AGN is preferentially associated with low-mass objects. However, when do we witness the onset of this trend? The answer to this question can be found in the right-hand plot of Figure  \ref{fig:idsvsmass} which clearly shows that FIR emitters almost entirely appear in relatively low-mass sources only in the local, $z\simlt 1$ universe. This phenomenon then loses much of its importance for redshifts $z\simgt 1$ and entirely disappears 
for $z\simgt 2$, whereby one finds that all radio-selected AGN have the same chances of also being FIR emitters independent of the stellar mass associated with their host galaxy.  

A trend which is very similar to what observed in the distributions of both radio luminosities and stellar masses of radio-selected AGN is also found in the distribution of the ages $\tau$ of their hosts, which again come from the work of Laigle et al. (2016). Indeed, in the left-hand panel of Figure \ref{fig:tau} it is possible to appreciate the systematic younger ages of FIR emitters (again represented with the red, dashed line) with respect to the total AGN population: the relative fraction of FIR emitters decreases for ages beyond $\sim 10^9$ yr, and very few of such sources are found for $\tau\simgt 10^{9.5}$ yr.
But again, as shown in the right-hand panel of Figure \ref{fig:tau}, this trend is mostly true in the local, $z\simlt 1$, universe, while it starts losing its importance at higher redshifts and by $z\simeq 2$ there are no appreciable differences between the ages of radio-AGN which are also FIR emitters and those of the whole radio-AGN population. 

As a caveat, we stress that the results presented in the top right-hand panels of Figures 2, 3 and 4 were obtained for sources up to $z=3$. However, we note that the McAlpine et al. (2013) data on which our selection method is based only extends to $z=2.5$. 
To make sure our results are not biased by an incorrect extrapolation to higher redshifts, we have then re-calculated all the previous quantities for sources in the redshift range $z=[2-2.5]$. These are represented in the top right-hand panels of Figures 2, 3 and 4 by the dashed histograms. As it is possible to appreciate, the distributions in the two $z=[2-3]$ and $z=[2-2.5]$ intervals are virtually identical. This, together with the fact that very little evolution is observed in the AGN luminosity function of McAlpine et al. (2013) in the whole  $z=[1.8-2.5]$  redshift range, gives us confidence that the extrapolation performed to select radio-emitting AGN for $z\ge 2.5$ is a sensible one.

Also, we made sure that contamination from star-forming galaxies to the AGN sample in the proximity of $P_{\rm cross}$ did not affect our results and conclusions. To this aim, we have recalculated all the quantities presented in Figures 2, 3 and 4 for a smaller AGN subset obtained by considering only those 444 sources with $P(z) > P^*(z)\equiv 2 \cdot P_{\rm cross}(z)$. The choice for such a new luminosity threshold ensures that the possible fraction of contaminants is now drastically reduced to a very negligible quantity at all redshifts (cfr the Appendix). While, by construction, this selection reduces the number of low-luminosity AGN and therefore of high redshift sources, no other difference is observed in any of the distributions presented in this Section. Therefore, we can safely conclude that possible contamination of the AGN sample at the lowest radio luminosities due to the presence of unremoved star-forming galaxies is not expected to affect any of our conclusions which can be summarised as:  {\it FIR emitters differentiate from the whole population of radio-selected AGN only in the local, $z\simlt 1$,  universe. At higher redshifts there is no difference between the properties of FIR-active and FIR-inactive sources and these two classes are indistinguishable one from the other.}


\section{FIR properties}

Magliocchetti et al. (2014) and (2016) have shown that FIR emission from the hosts of radio-active AGN is entirely to be attributed to star-forming processes within the galaxy itself. This is also true in the present case. In fact, as Figure \ref{fig:q100_160} clearly illustrates, except for a very few outliers, the FIR colours of radio-selected AGN of all luminosities and at all redshifts all lie on the curves identified by the spectral energy distributions (SED) of  two standard star-forming galaxies: M82 (dashed line) and Arp220 (dotted line). 
Despite this finding, in the majority of cases the flux ratios $\rm q_{100}=log_{10}\left[ F_{100 \mu m}/F_{1.4 GHz}\right ]$ and $\rm q_{160}=log_{10}\left[F_{160 \mu m}/F_{1.4 GHz}\right ] $ of these sources indicate an excess of radio activity with respect to that produced by 'pure' star-forming processes, most likely to be attributed to the presence of a central, radio-emitting, AGN. This is seen in both panels of Figure \ref{fig:q100} which illustrate the distribution of the quantities q$_{100}$ (left-hand plot) and q$_{160}$ (right-hand plot) as a function of redshift for radio-selected AGN of different radio luminosities, once again compared with the SEDs obtained  for M82 and Arp220. 

\begin{figure}
\includegraphics[scale=0.4]{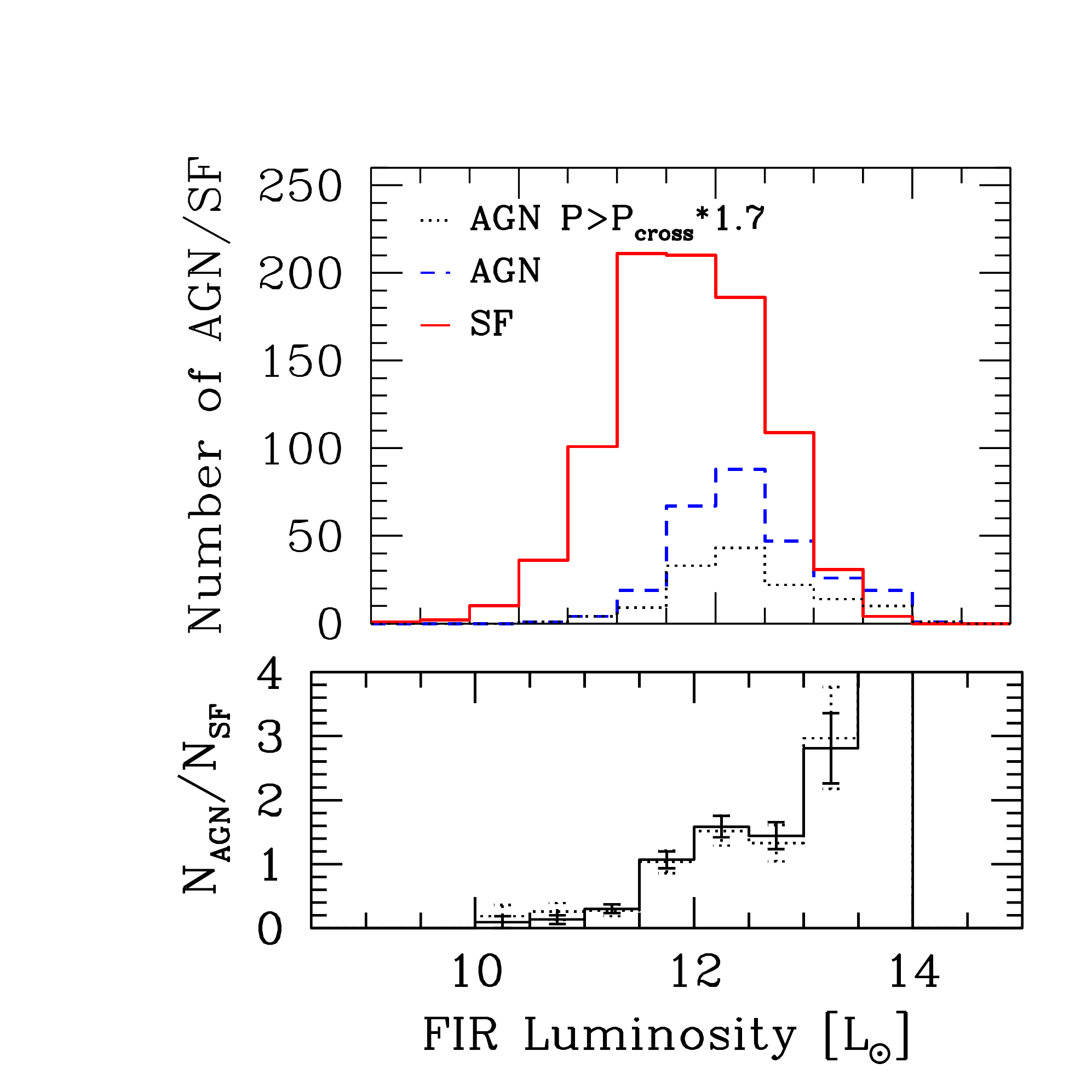}
\caption{Distribution of FIR luminosities for radio-selected AGN (blue, dashed line) and radio-selected star-forming galaxies (red, solid line) with a counterpart in the PEP maps. The dotted histogram indicates the luminosity distribution of radio-selected AGN with radio luminosities ${\rm Log}_{10} P={\rm Log}_{10} P_{\rm cross}+0.2$. The bottom panel represents the ratio between these three quantities: AGN/SF (solid line) and AGN with  ${\rm Log}_{10} P={\rm Log}_{10} P_{\rm cross}+0.2$ over SF (dotted line). These are all appropriately rescaled for the total number of sources in the considered catalogues. Errorbars correspond to $1 \sigma$ Poissonian estimates.
\label{fig:hist_lir}}
\end{figure}


Some interesting insights on the mechanisms which regulate star-formation in radio-selected AGN can be obtained from a comparison of the FIR properties of  this class of objects with those exhibited by the population of star-forming galaxies, where once again this latter class has been  selected from the Bondi et al. (2008) catalogue by following the method highlighted in \S 3.  To this aim, we have calculated the bolometric luminosities L$_{\rm FIR}$ of both populations by integrating over the whole 8$\mu$m-1000$\mu$m range chosen templates of star-forming galaxies which best fitted the FIR data and subsequently extrapolated their star formation rates (SFR)  according to the standard relation (Kennicut 1998, which holds for a Salpeter IMF and for stellar masses in the range $\sim 0.1-100$ M$_\odot$): SFR [M$_\odot$ yr$^{-1}$]=$1.8 \cdot 10^{-10}$ L$_{\rm FIR}$/L$_{\odot}$.

The distribution of L$_{\rm FIR}$ for both classes of sources is presented in Figure \ref{fig:hist_lir}, whereby star-forming galaxies are represented by the solid red histogram, while the blue dashed one is for AGN. First of all, our data clearly show that radio-active AGN are extremely luminous sources. Their distribution peaks at around 10$^{12.5}$ L$_\odot$ and presents a tail of sources as bright as $\sim 10^{14}$ L$_\odot$. Also, radio-selected AGN are on average brighter than star-forming galaxies selected from the same radio sample. Indeed, these latter objects present a distribution which peaks at lower, L$_{\rm FIR}=\left [ 10^{11}-10^{12} \right]$ L$_\odot$, luminosities and there are practically no sources with L$_{\rm FIR}$ beyond $10^{13}$ L$_\odot$. This is better seen in the bottom panel of Figure \ref{fig:hist_lir} which shows the relative weight of the two populations of radio-selected sources as a function of FIR luminosity. The fraction of AGN drastically increases for L$_{\rm FIR}\simgt 10^{11.5}$ L$_\odot$, and they become the dominant population in sources brighter than $\sim 10^{13}$ L$_\odot$.  We note that  such a finding is robust with respect to possible contaminations of the AGN sample. In fact, if we concentrate on a smaller subset obtained by only considering radio-selected AGN with luminosities brighter than 
$P(z)=2 \cdot P_{\rm cross}(z)$, we obtain the very same trend (cfr dotted histogram in the bottom panel of Figure 7), despite having only used roughly  half of the available sources.

The above result implies that the presence of a radio-active AGN within a galaxy not only does not inhibit its star-formation activity but, on the contrary, that such an activity is probably favoured by the very same presence of the central AGN. 
In other words, what we are witnessing is the likely presence of positive feedback, possibly due to the winds produced by the radio-active AGN which boost star-formation within its host. This issue will be better discussed in \S 5.

\begin{figure*}
\includegraphics[scale=0.28]{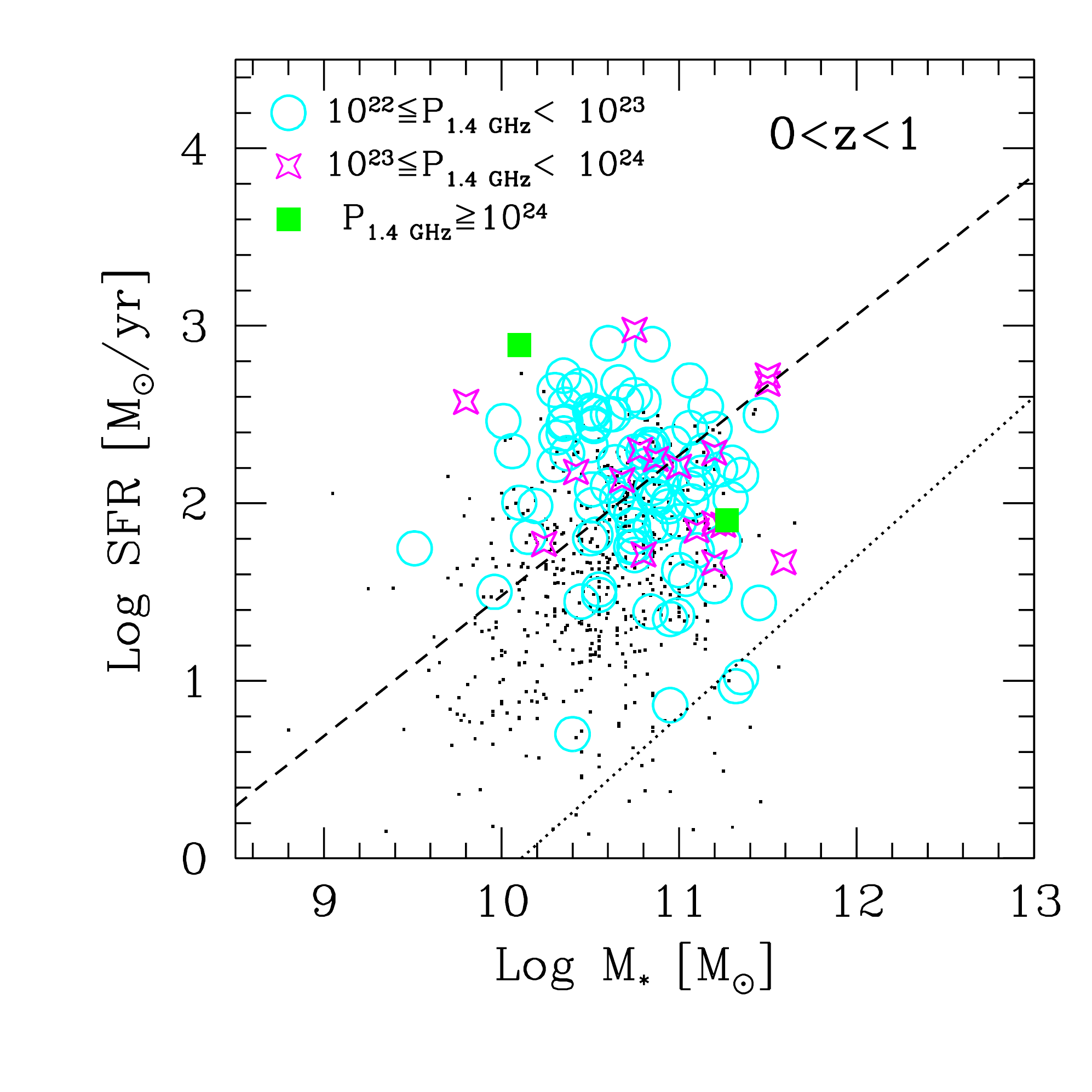}
\includegraphics[scale=0.28]{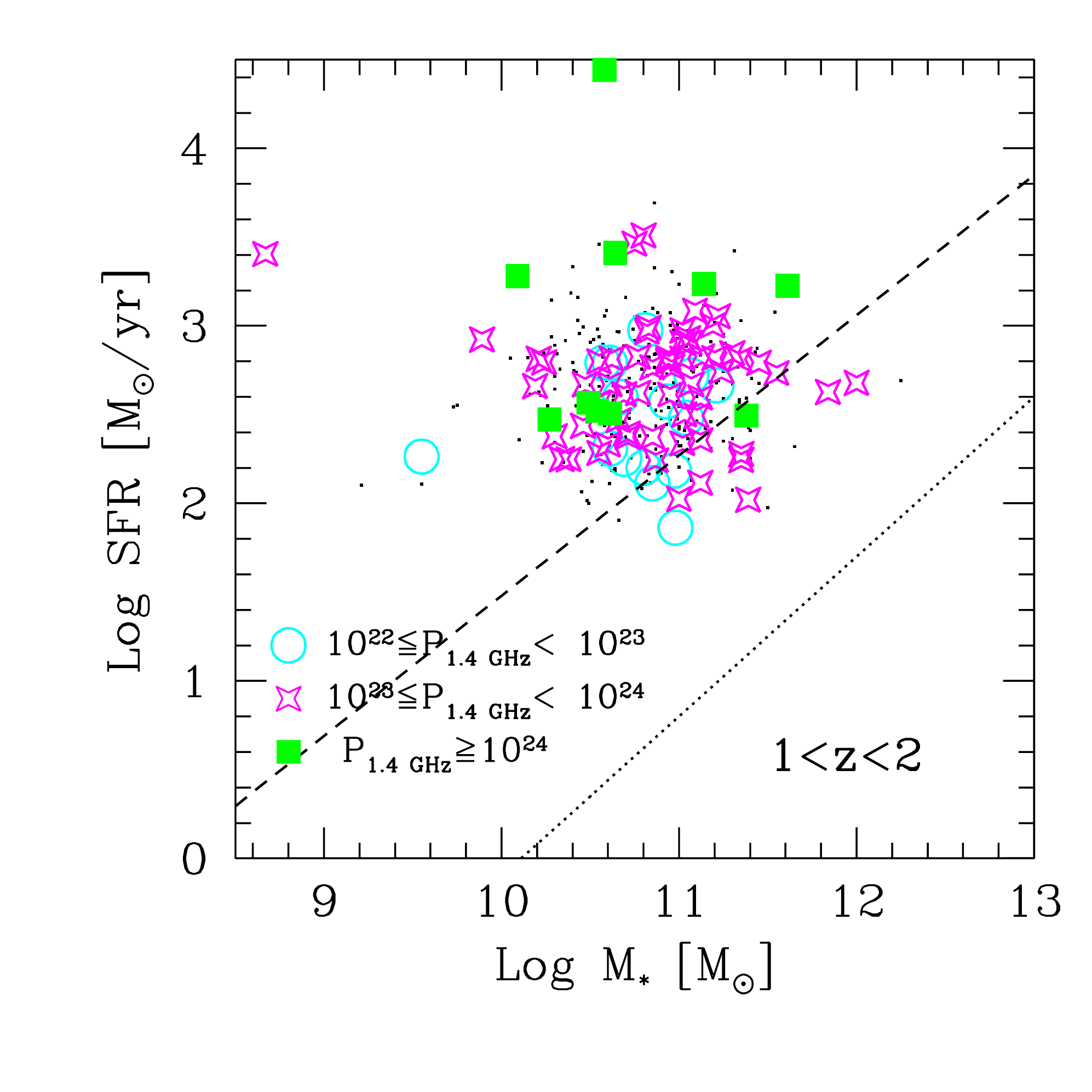}
\includegraphics[scale=0.28]{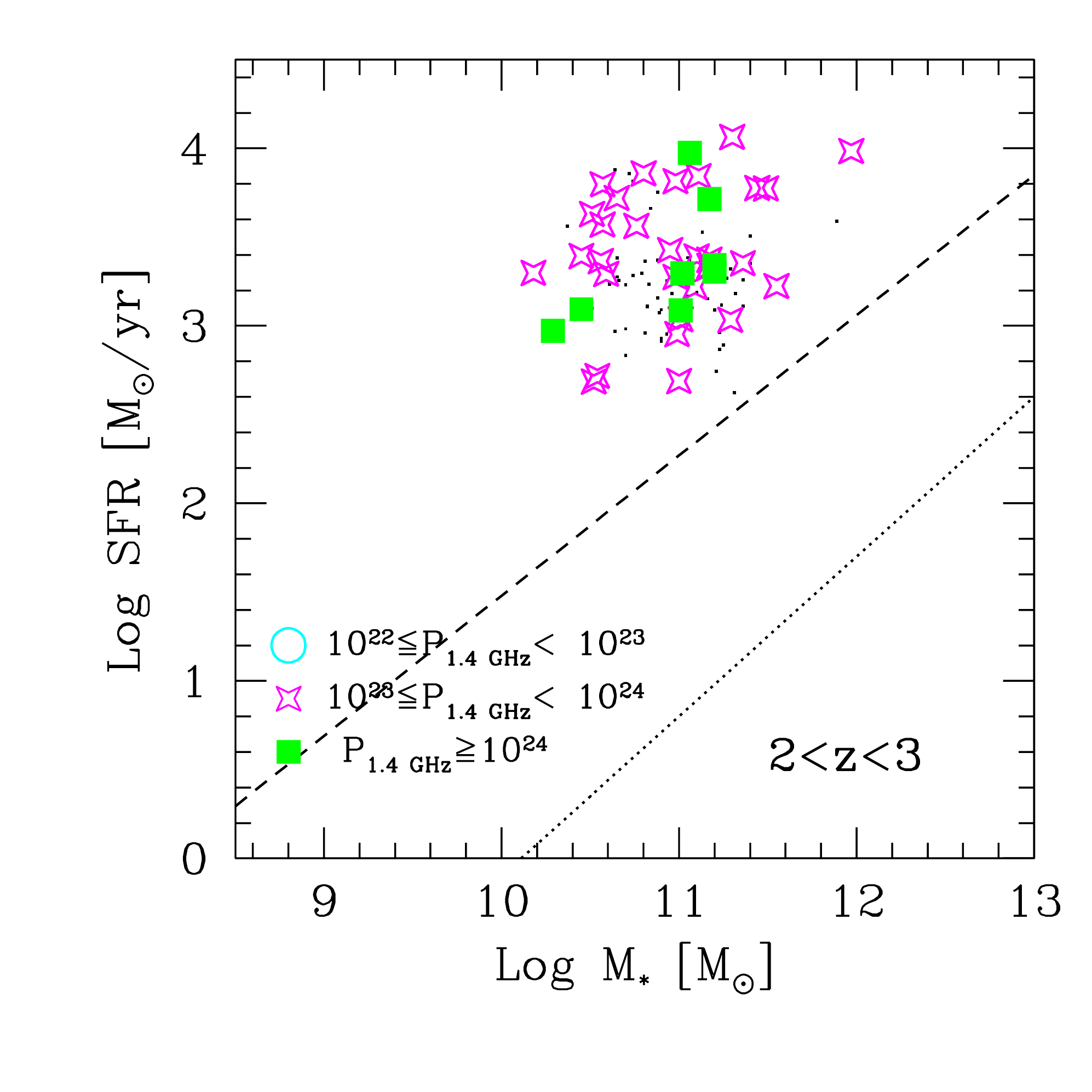}
\caption{Star formation rates as a function of stellar mass for radio-selected AGN in three different redshift intervals. Sources are colour-coded according to their radio luminosity (expressed in [W Hz$^{-1}$ sr$^{-1}$] units). Black dots indicate the distribution observed for radio-selected star-forming galaxies of all radio luminosities. The dashed lines indicate the relation obtained for main sequence galaxies at z $\sim 2$ by Rodighiero et al. (2011), while the dotted lines that derived for local galaxies by Brinchmann et al. (2004) and Peng et al. (2010).
\label{fig:sfrz}}
\end{figure*}

\begin{figure*}
\includegraphics[scale=0.28]{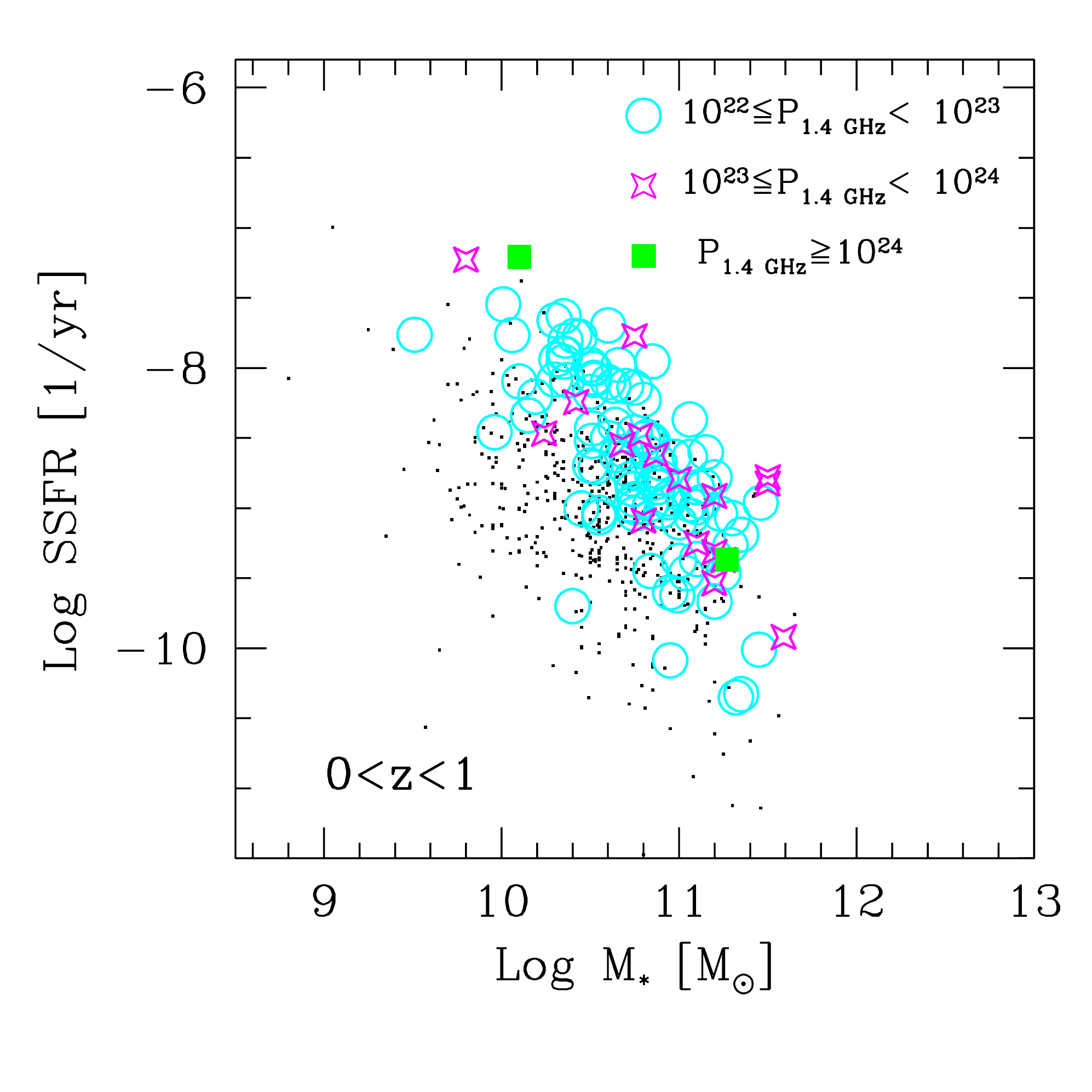}
\includegraphics[scale=0.28]{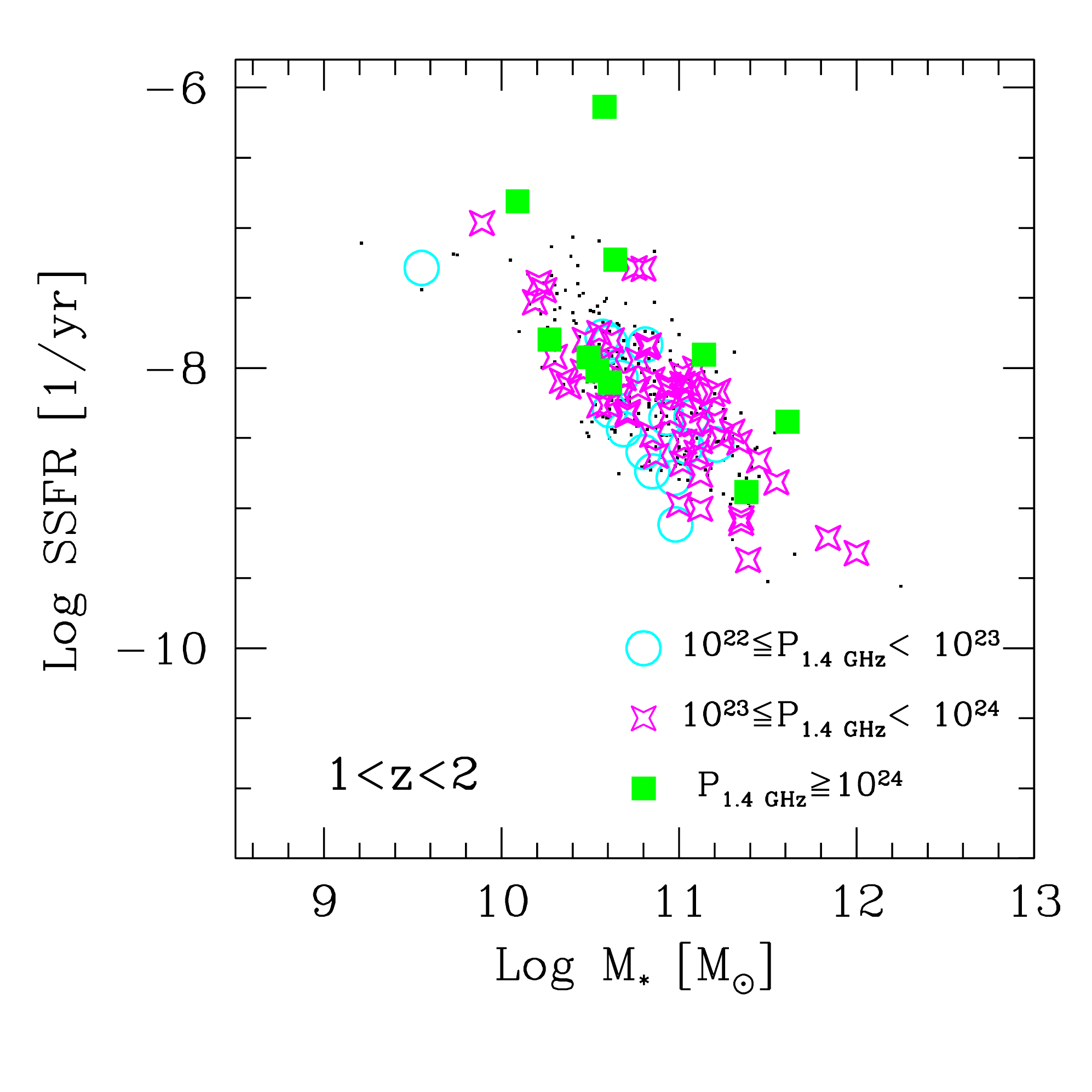}
\includegraphics[scale=0.28]{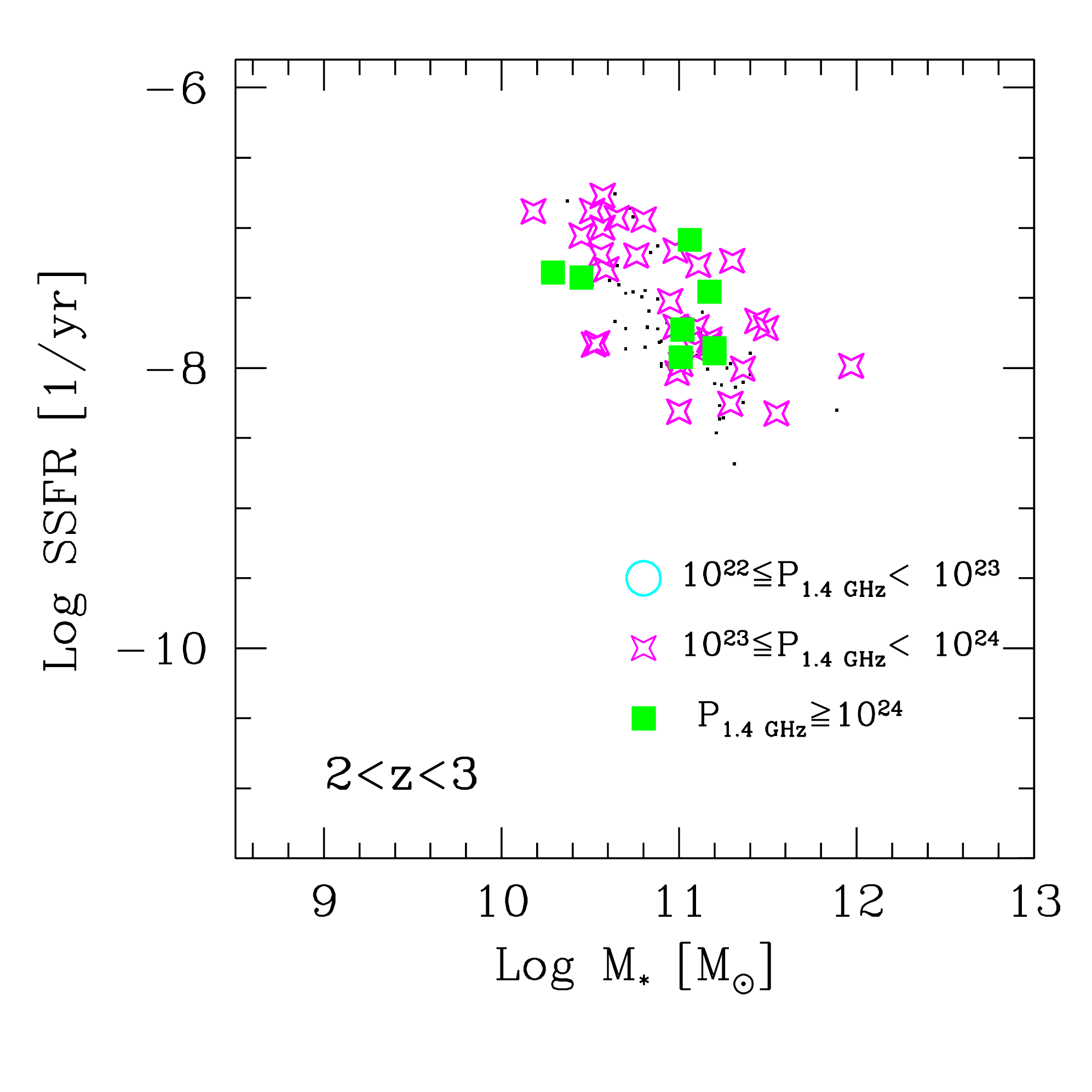}
\caption{Specific star formation rates as a function of stellar mass for radio-selected AGN in three different redshift intervals. Sources are colour-coded according to their radio luminosity (expressed in [W Hz$^{-1}$ sr$^{-1}$] units). Black dots indicate the distribution observed for radio-selected star-forming galaxies of all radio luminosities.
\label{fig:ssfrz}}
\end{figure*}

Some more differences between the two populations of radio-selected AGN and star-forming galaxies can be appreciated by investigating the distributions of their SFR and Specific Star Formation Rates (SSFR, defined as SFR over stellar mass) as a function of stellar mass. This is done in Figures \ref{fig:sfrz} and \ref{fig:ssfrz}, where the three panels within each Figure refer to different redshift ranges. In both Figures, star-forming galaxies of all radio luminosities are represented as black dots, while FIR-emitting AGN are colour coded according to their radio power. In order to guide the eye, the dashed and dotted lines in Figure \ref{fig:sfrz} respectively indicate the relation obtained for main sequence galaxies at $z\sim 2$ by Rodighiero et al. (2011) and  that derived for local galaxies by Brinchmann et al. (2004) and Peng et al. (2010). What is clearly visible from both Figures is that, independent of their radio power, FIR-active AGN and star-forming galaxies are indistinguishable from each other at all redshifts beyond $z\simeq 1$. Indeed, they present the very same distribution of stellar masses, star-forming rates and consequently of specific star-forming rates. 
In the local universe though this similarity breaks down and the two populations exhibit different properties: AGN tend to be more massive and, most of all, more FIR-bright than star-forming galaxies. The star-formation rates of this latter population can be as low as $\sim 0$ and do not extend further than $\sim 10^2$ M$_\odot$/yr. On the other hand, irrespective of their radio luminosity, FIR-emitting AGN are basically only found above SFRs $\simeq 30$ M$_\odot$/yr and can present SFRs as high as $10^3$ M$_\odot$/yr. Indeed, radio-selected star-forming galaxies and AGN at $z\simlt 1$ occupy two well defined portions of the SSFR-Mass plane (cfr left-hand panel of Figure \ref{fig:ssfrz}), with very little superposition between the two populations.

These results are in agreement and extend the findings of the former sections which indicate that for redshifts above $\sim 1$ there is very little or even no difference between radio-selected AGN and the sub-population of those which are also active at FIR wavelengths. Here we have seen that, at the same high redshifts, there is also no difference between FIR-active, radio-selected AGN and FIR-active, radio-selected star-forming galaxies. 
In all cases, differences only appear in the local universe: radio-active AGN progressively start becoming FIR-quiet, the more as the higher are their radio luminosities and the mass and age of their hosts, so that FIR activity remains preferentially locked in galaxies which are smaller and younger than the average parent population (cfr \S3).
At the same time, AGN are found associated with hosts of masses which are also higher than those characterising star-forming galaxies, and the few AGN which still emit at FIR wavelengths in the local universe on average exhibit much higher FIR luminosities (and consequently star-forming rates) than sources undergoing a pure process of stellar formation (cfr left-hand panels of Figures \ref{fig:sfrz} and \ref{fig:ssfrz}).


\section{Radio-selected AGN of different nature}

\begin{table}
\begin{center}
\caption{Number of radio-selected AGN with fluxes F$_{\rm 1.4 GHz}\ge 0.06$ mJy as taken from the COSMOS-VLA survey.  The first row refers to the whole sample, the second one to those which also show AGN emission in the MIR band, while the third row is for sources which are also identified as AGN in the X-ray band.
N$_{\rm TOT}$ refers to the total number of AGN, while  
N$_{\rm FIR}$ indicates the number of objects which are also detected at FIR wavelengths. The percentage symbols indicate the percentage of radio-selected AGN which are also detected at the different wavebands considered in our work. In the second column we report those for all sources, while the fourth column indicates those for radio-selected AGN which also emit in the FIR.}
\begin{tabular}{llllll}
  & N$_{\rm TOT}$& \% & N$_{\rm FIR}$& \%\\
\hline
\hline
Parent sample& 704 & - & 272&- &\\
MIR detected & 141& 20.0\%&83&30.5\%\\
X-ray detected& 182& 25.8\%&81&29.8\%\\
\hline
\hline

\end{tabular}
\end{center}
\end{table}

\begin{figure*}
\includegraphics[scale=0.35]{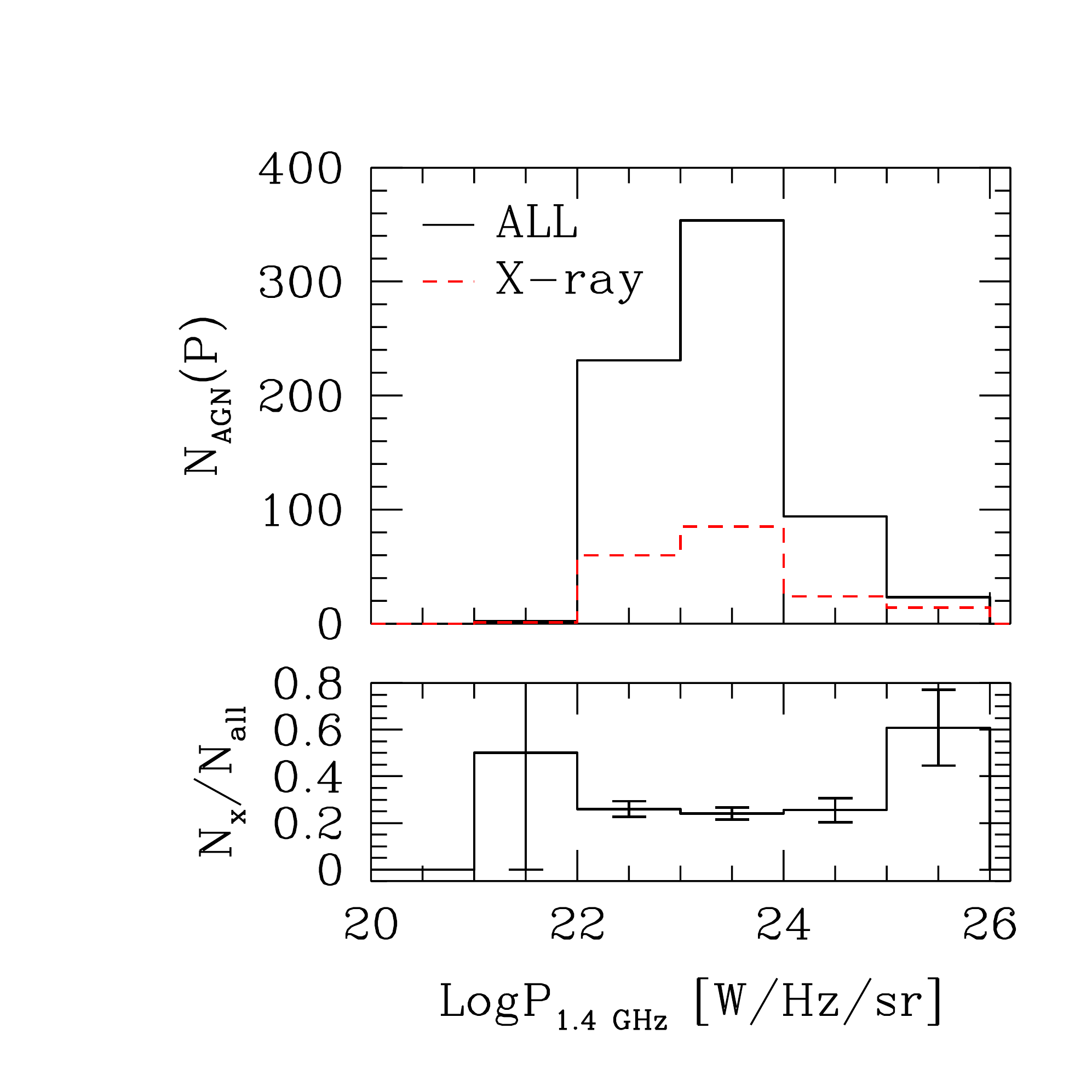}
\includegraphics[scale=0.35]{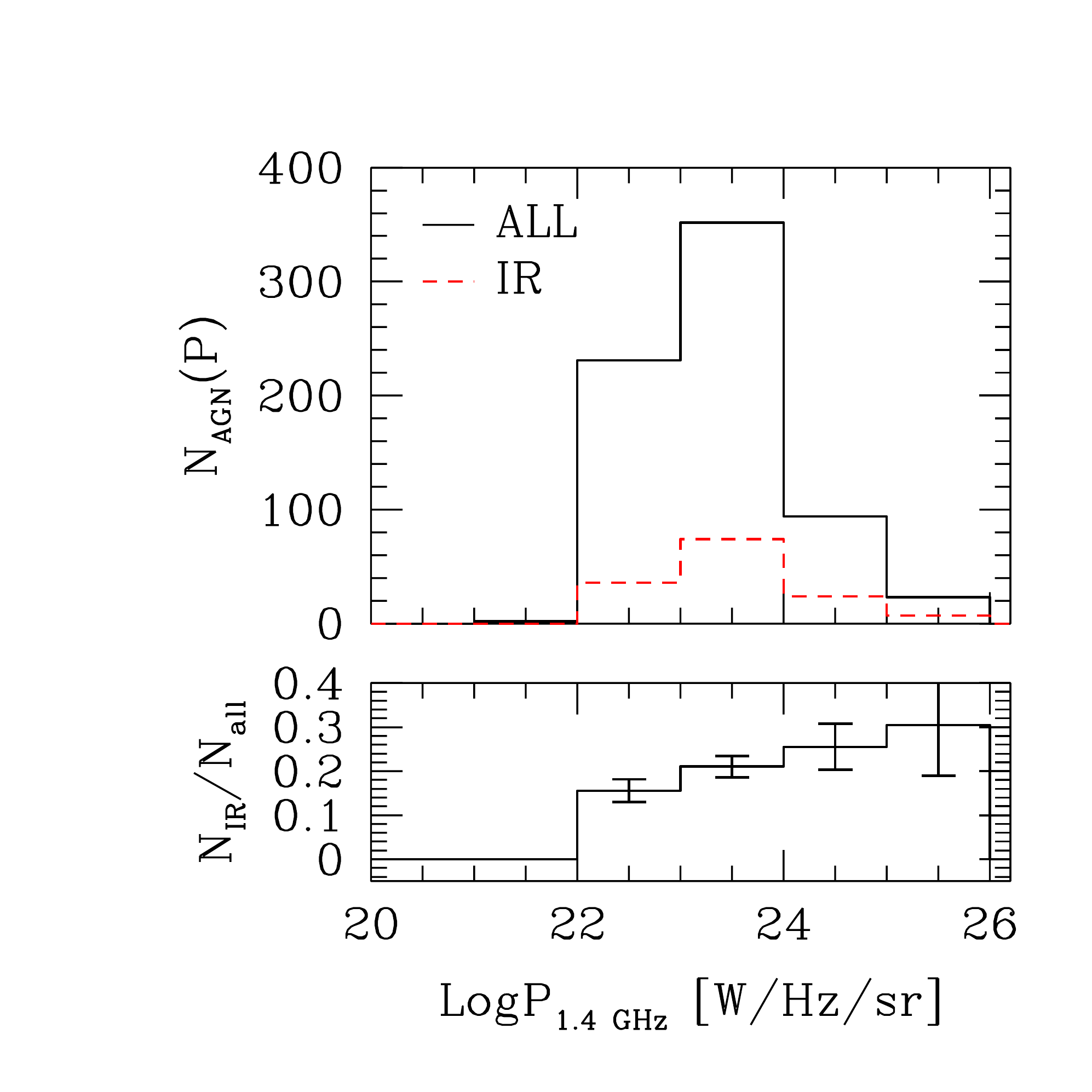}
\caption{Distribution of radio powers for F$_{\rm 1.4 GHz}\ge 0.06$ mJy COSMOS-VLA radio-selected AGN of different types. The dashed line in the  left-hand panel represents sources which are also classified as AGN in the X-ray, that in the right-hand panel those which present AGN emission in the MIR band.
The solid lines are for all radio-selected AGN.  The bottom panels highlight the ratio between the two quantities. Errorbars correspond to $1 \sigma$ Poissonian estimates.
\label{fig:Phist_type}}
\end{figure*}

\begin{figure*}
\includegraphics[scale=0.35]{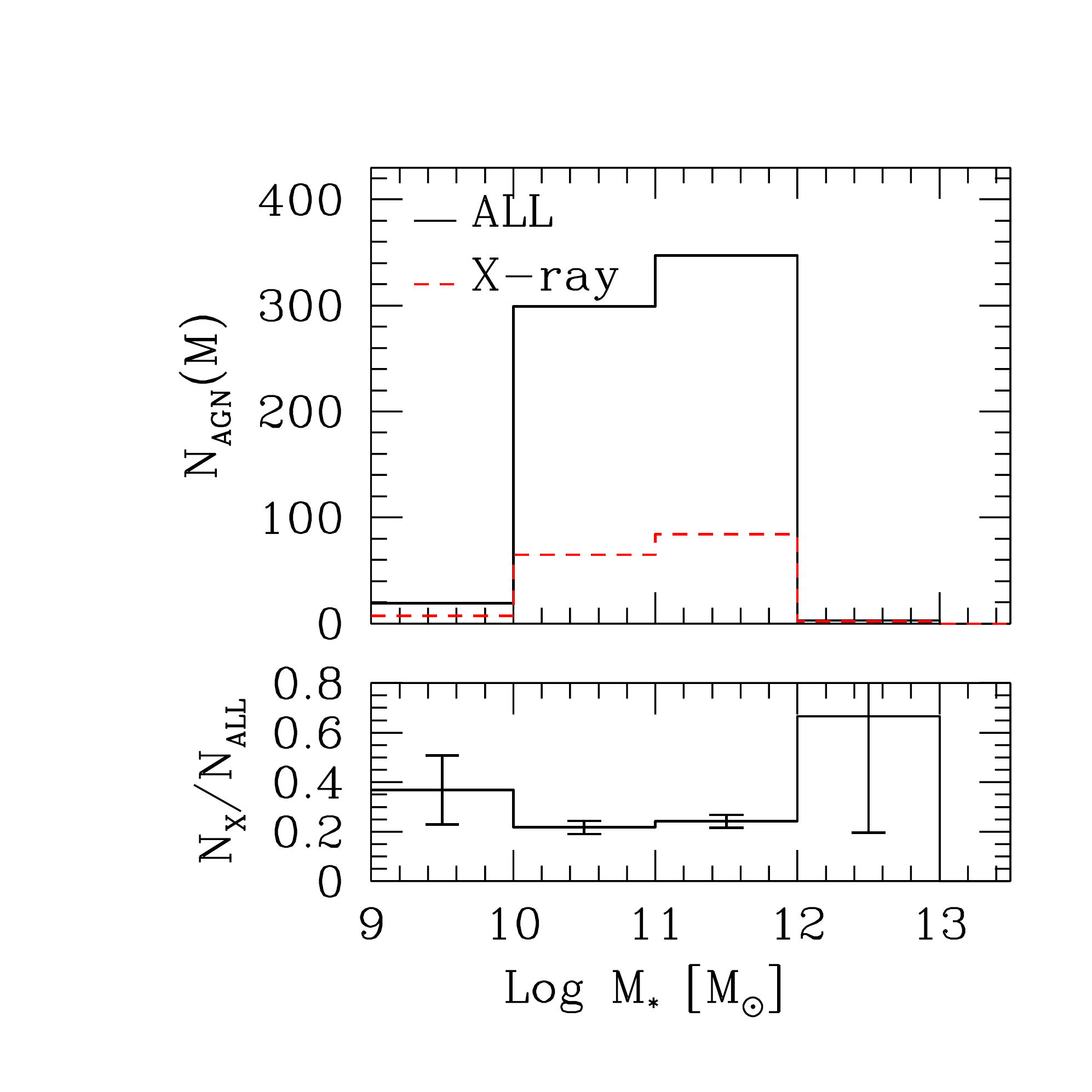}
\includegraphics[scale=0.35]{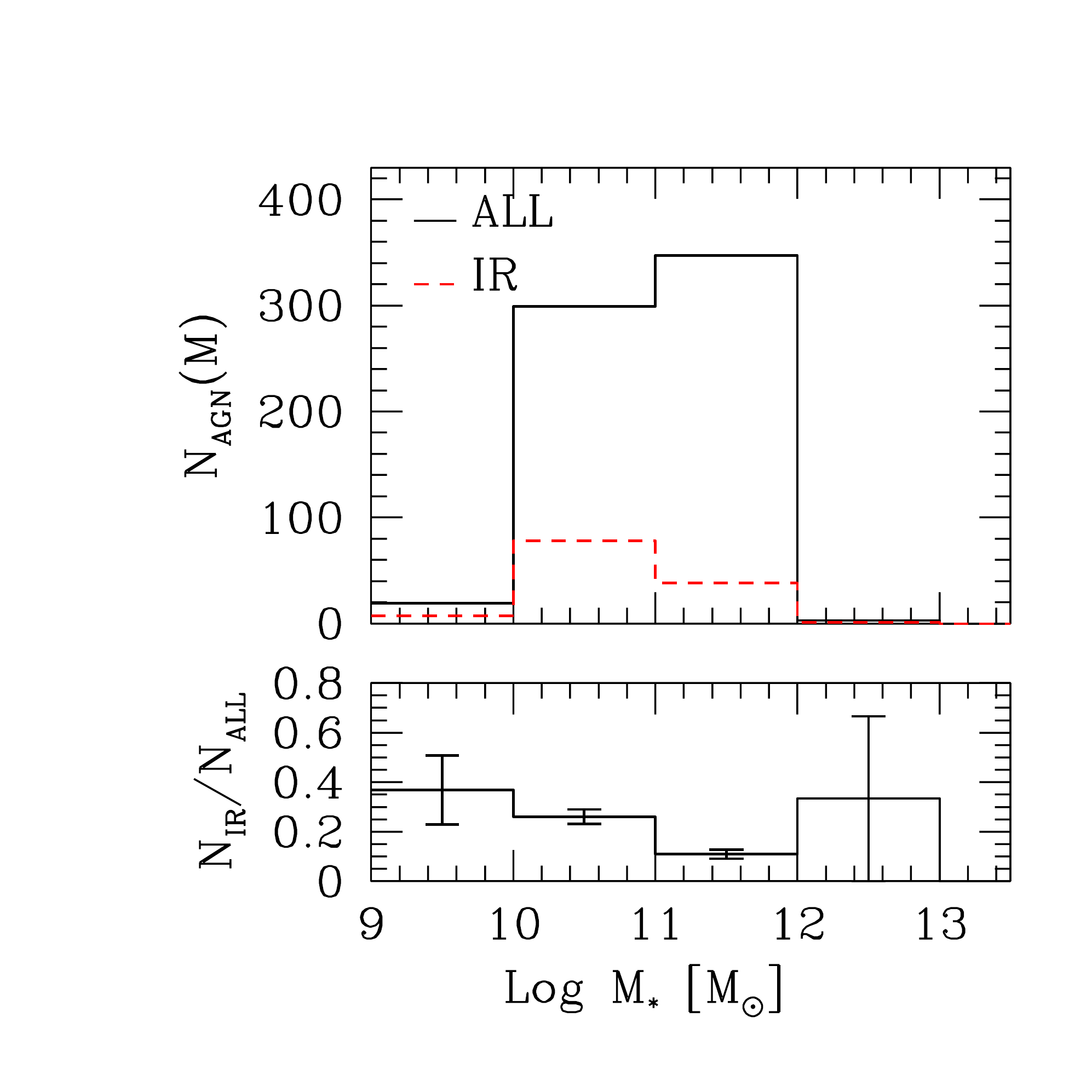}
\caption{Distribution of stellar masses for F$_{\rm 1.4 GHz}\ge 0.06$ mJy COSMOS-VLA radio-selected AGN of different types. The dashed line in the  left-hand panel represents sources which are also classified as AGN in the X-ray, while that in the right-hand panel those which present AGN emission in the MIR band.
The solid lines are for all radio-selected AGN.  The bottom panels highlight the ratio between the two quantities. Errorbars correspond to $1 \sigma$ Poissonian estimates.
\label{fig:masshist}}
\end{figure*}

\begin{figure*}
\includegraphics[scale=0.35]{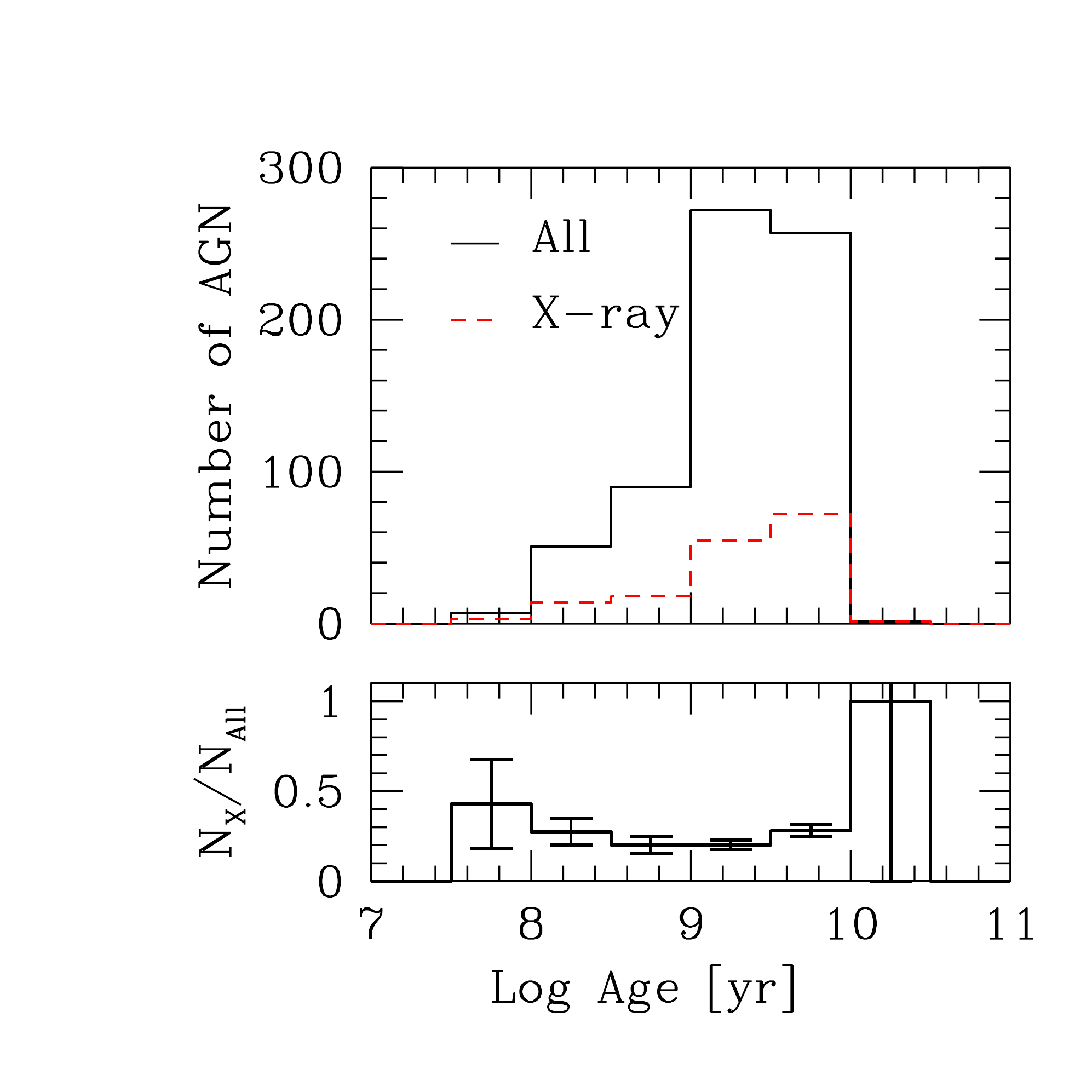}
\includegraphics[scale=0.35]{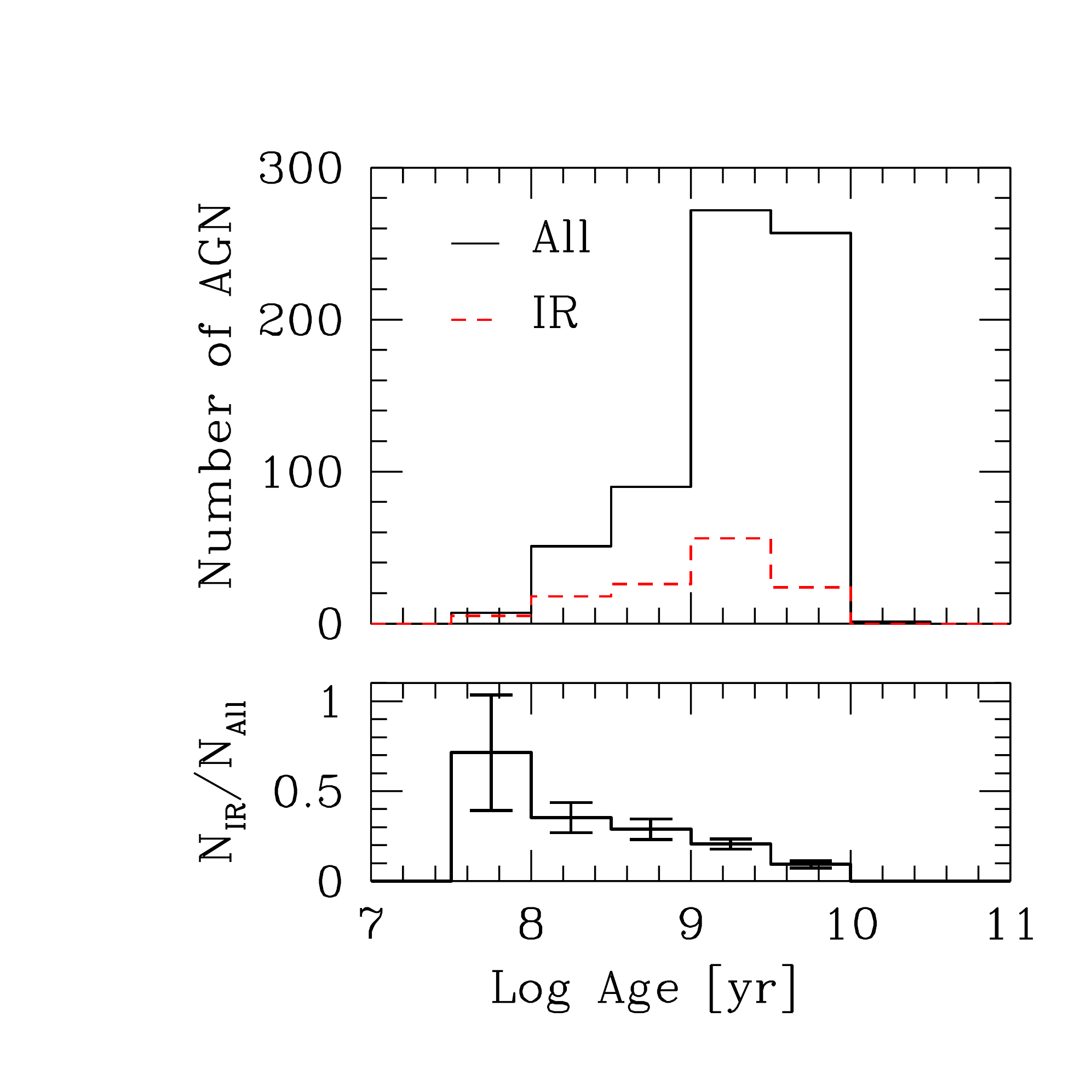}
\caption{Distribution of ages $\tau$ for the hosts of F$_{\rm 1.4 GHz}\ge 0.06$ mJy COSMOS-VLA radio-selected AGN of different types. The dashed line in the  left-hand panel represents sources which are also classified as AGN in the X-ray, while that in the right-hand panel those which present AGN emission in the MIR band.
The solid lines are for all radio-selected AGN.  The bottom panels highlight the ratio between the two quantities. Errorbars correspond to $1 \sigma$ Poissonian estimates.
\label{fig:tauhist}}
\end{figure*}

Galaxies in the COSMOS catalogue provided by Laigle et al. (2016) are also flagged according to whether they present signatures of AGN activity in their spectra or  SEDs. This is then also true for the radio-selected AGN considered in the present work. We therefore looked for flags in order to determine whether our sources were also identified as AGN in the X-ray or MIR bands. We stress that we had to limit our analysis only to these two sub-populations as, unlike the products of X-ray and MIR selection,  optically-identified AGN do not belong to a homogeneous sample, therefore no statistical information could be drawn from these objects. 

X-ray information comes from the {\it Chandra COSMOS-Legacy Survey} (Civano et al. 2016; Marchesi et al. 2016) which includes X-ray sources down to a flux limit $f_X=2\cdot 10^{-16}$ erg s$^{-1}$ cm$^{-2}$ in the 0.5-2 KeV band. 
MIR selection is instead based on the power-law behaviour in the Mid-Infrared/{\it Spitzer}-IRAC bands of the SEDs of the considered sources (Donley et al. 2012; Chang et al. 2016 in preparation).

The results of our cross-match are summarised in Table 1. Between 20\% and 26\% of radio-emitting AGN are also identified as AGN respectively in the MIR or X-ray band. This figure rises up to $\sim$30\% in the case of MIR AGN which are also detected at FIR wavelengths, while it basically stays constant in the case of X-ray-emitting AGN. These findings imply a mild preference for radio-selected AGN which also emit in the MIR to appear in sources associated with stellar production, while X-ray emission of AGN origin seems to be roughly independent of the FIR activity of its host galaxy.

With the above information at hand, we can investigate the very nature of radio-active AGN and determine whether there is any substantial difference between those that are only active at radio wavelengths and those which also emit in other parts of the spectrum. 

First of all, as shown in Figure \ref{fig:Phist_type} and with the possible exception of the highest luminosity regime probed by our analysis,  there is a hint that,  while the chances for a radio-active AGN to also be an X-ray emitter seem to be independent of its radio luminosity,  this is not true for AGN which instead also emit in the MIR band, as  the probability for MIR emission seems to be monotonically enhanced in radio-selected AGN of higher and higher radio luminosity.

A similar behaviour is also observed in the distribution of the stellar masses of the hosts of radio-selected AGN (cfr Figure \ref{fig:masshist}). In fact, even in this case we observe a substantial consistency of the relative fraction of X-ray emitters with stellar mass, while there is a clear trend for radio-active AGN which also emit in the MIR to preferentially reside in galaxies of smaller mass. At the same time, radio-selected AGN which are also X-ray emitters show the same distribution of ages $\tau$ as the parent radio-AGN population, while those radio-AGN which also emit in the MIR are found to be associated with systematically younger galaxies (cfr Figure \ref{fig:tauhist}).

\begin{figure}
\includegraphics[scale=0.40]{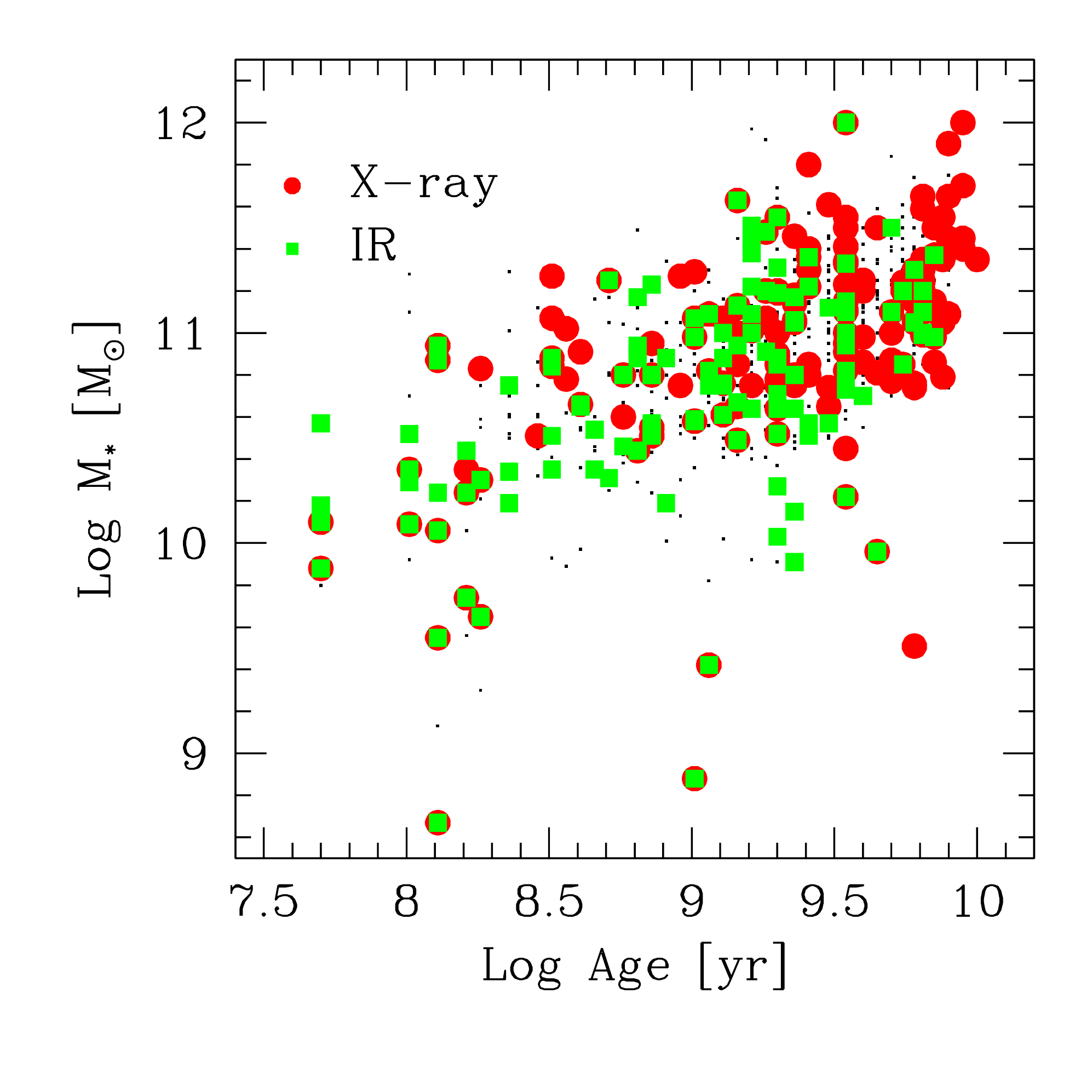}
\caption{Stellar masses as a function of ages for those galaxies host of a radio-selected AGN.  Filled (red) circles identify AGN which also emit in the X-ray, while (green) squares represent AGN which show signs for AGN activity also in the MIR band. The black dots represent the whole parent radio-AGN population.
\label{fig:massvsage}}
\end{figure}

The above comparisons highlight an interesting fact: the sub-population of radio-selected AGN which also emit in the X-ray constitutes a class of sources which is indistinguishable from its parent population in terms of radio luminosity  and also of stellar masses and ages of their host galaxies. There is neither a preferential level of radio activity of the central black hole nor a preferential sub-galaxy environment which can determine whether the radio-active AGN will also emit in the X-ray band or not. On the contrary, MIR emission of AGN origin in radio-selected AGN seems to be favoured in sources which are radio-bright and hosted by galaxies which are relatively small and young.

This effect is better visible in Figure \ref{fig:massvsage} which shows the distribution of stellar masses and ages for the hosts of radio-selected AGN which are also active in the  MIR (green squares) and X-ray (red circles) bands: AGN which are also active in the X-ray are systematically more massive and older than the other class of sources. We can therefore envisage an evolutionary track for radio-active AGN which associates MIR  emission of AGN origin mainly with the early stages of the life-time of the radio source, when the host galaxy was young and fewer stars were already in place. Note that this also agrees with our finding (cfr Table 1) of an enhanced fraction of MIR emitters in radio-selected AGN which cohabit with intense episodes of stellar formation. On the other hand, X-ray emission in radio-selected AGN  seems to be mainly associated with a later stage of the evolution of these sources, whereby their hosts have built most of their stellar mass and are on average relatively old. This is also true for radio-selected AGN which only show emission at radio wavelengths, as these two latter classes of sources seem to be indistinguishable from each other.

\begin{figure*}
\includegraphics[scale=0.35]{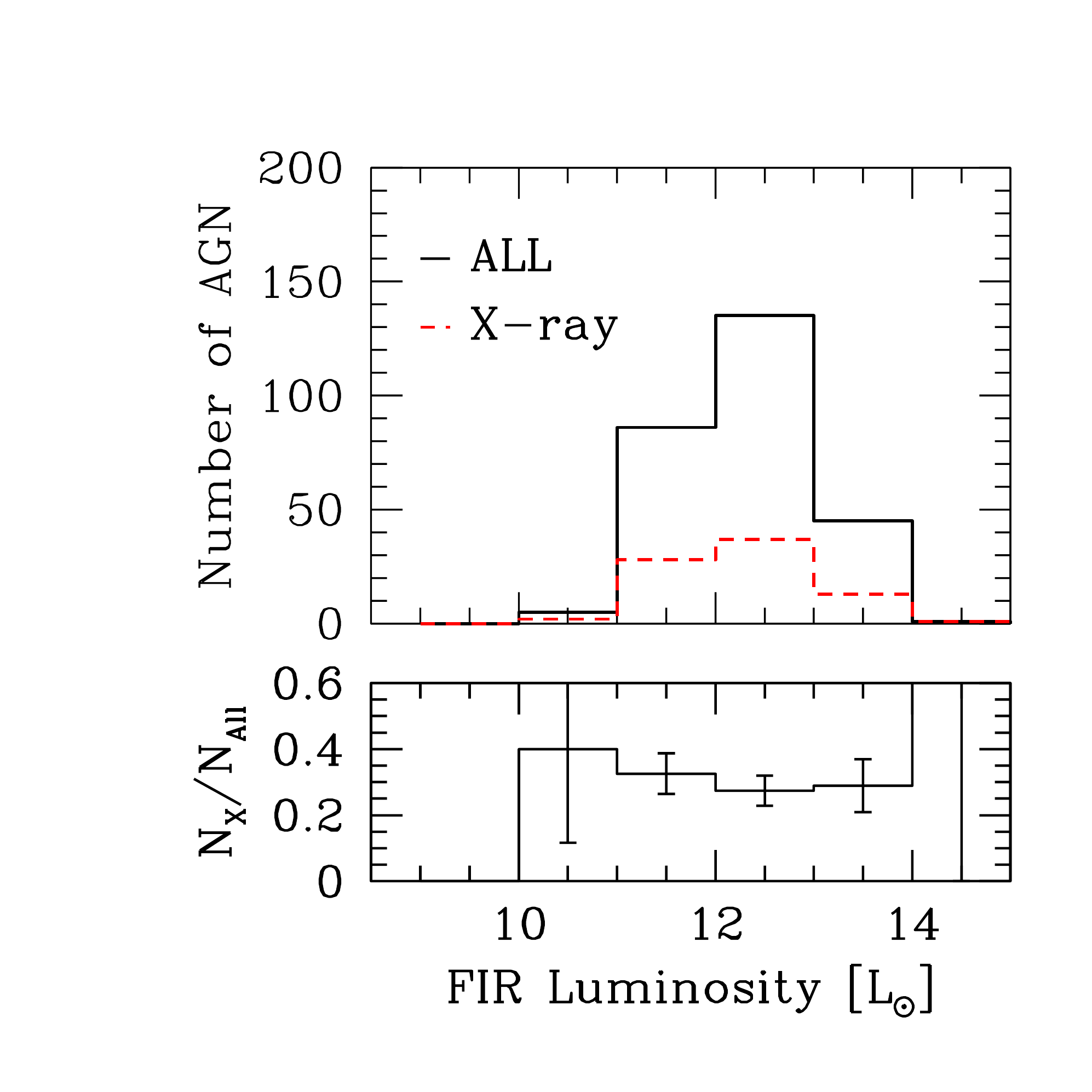}
\includegraphics[scale=0.35]{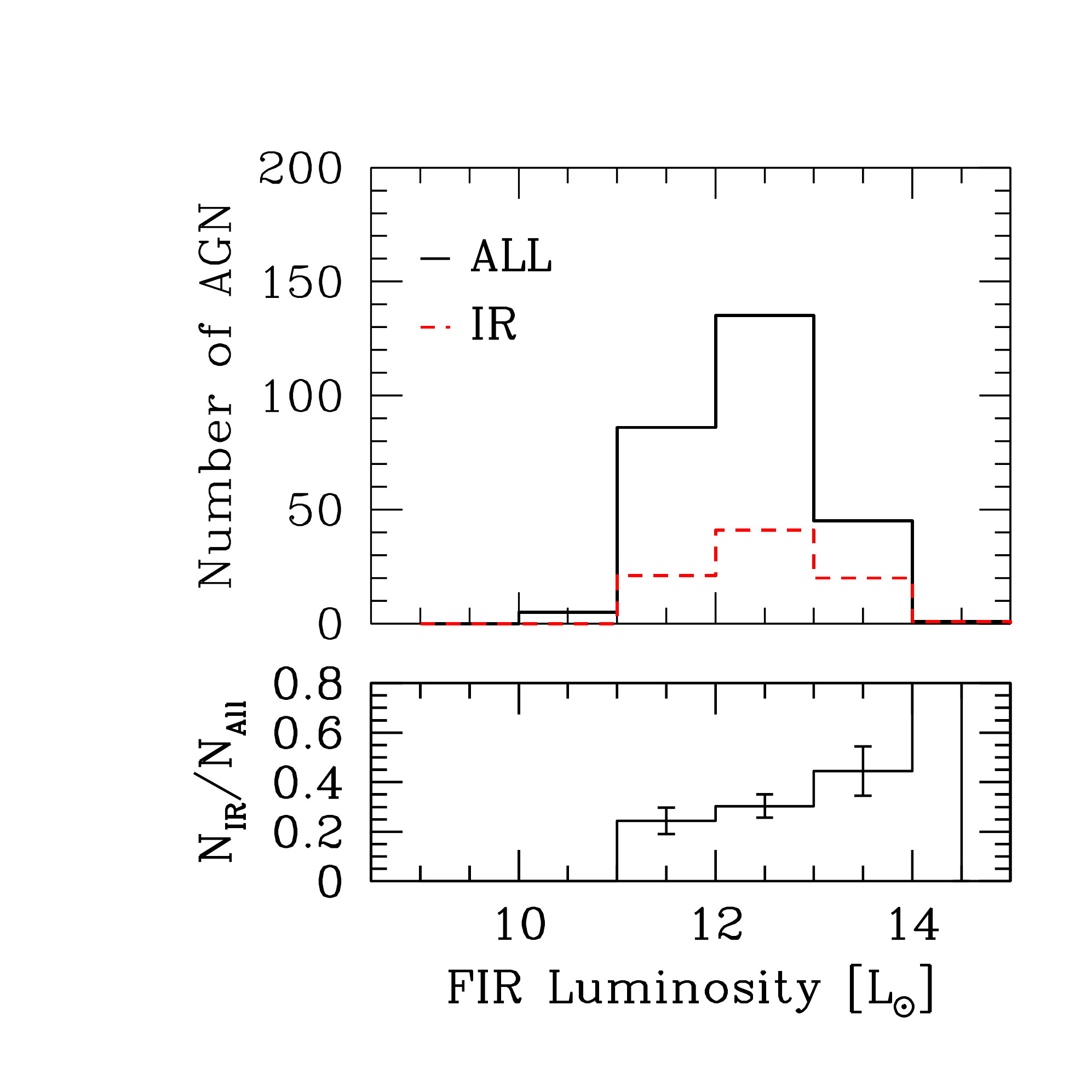}
\caption{Distribution of FIR luminosities for F$_{\rm 1.4 GHz}\ge 0.06$ mJy COSMOS-VLA radio-selected AGN of different types. The dashed line in the  left-hand panel represents sources which are also classified as AGN in the X-ray, while that in the right-hand panel those which present AGN emission in the MIR band.
The solid lines are for all radio-selected AGN which also present FIR emission.  The bottom panels highlight the ratio between the two quantities. Errorbars correspond to $1 \sigma$ Poissonian estimates.
\label{fig:lumhist_type}}
\end{figure*}

Amongst the pieces of information that can be gathered from Table 1, we also find that the majority of radio-active AGN which also emit in the MIR band is associated with FIR activity. Indeed, 83 out of 141 sources (corresponding to 59\% of the subsample) are detected in the COSMOS-{\it Herschel} maps, while only 44\% of radio-AGN with emission in the X-ray are instead found to be associated with ongoing star-formation within the host galaxy. 
This finding further confirms the conclusions previously reached:  MIR emission is mainly favoured in hosts which are younger than those hosting AGN which are only active in the radio or in the radio+X-ray bands. 
Indeed, as clearly shown in Figure \ref{fig:lumhist_type} which reports the distributions of FIR luminosities for these two sub-classes of AGN as compared with the distribution observed for the whole population of radio-selected AGN which show up in the {\it Herschel} maps, while the fraction of X-ray AGN stays constant for increasing FIR luminosities, this is found to monotonically increase in the case of AGN which also emit in the MIR band.



\section{conclusions}
By making use of the recent catalogue of redshifts produced by Laigle et al. (2016)  for galaxies belonging to the COSMOS field, we have performed a thorough analysis of the population 
of 1.4 GHz-selected galaxies on the same COSMOS area and on their Far-Infrared properties. This can be considered as a completion of the work presented in Magliocchetti et al. (2014). 

About 90\% of the sources from the VLA-COSMOS survey (Bondi et al. 2008) are found to have a counterpart in the Laigle et al. (2016) catalogue. 
These objects have then been sub-divided into radio-active AGN and radio-emitting star-forming galaxies solely on the basis of their radio luminosity. Out of 2123 radio sources endowed with a redshift estimate, 
704 (corresponding to $\sim 33$\% of the parent population) are AGN and the remaining star-forming galaxies. 
By then looking for FIR counterparts on the {\it Herschel}-PEP (Lutz et al. 2011) maps, we found that 272 of such radio-emitting AGN are also FIR emitters. The redshift distribution of the sub-class of FIR emitters mirrors that of the whole parent population of radio-active AGN and features a prominent peak at $z\simeq 1$ and a broad tail which extends up to $z\simeq 4$. 

The main conclusions that can be drawn from our analysis can be summarised as follows:
\begin{enumerate}
\item FIR emitters amongst radio-active AGN are preferentially found at low, LogP$_{1.4 \rm GHz}\simlt 10^{23}$ [W Hz$^{-1}$ sr$^{-1}$], radio luminosities. However, this is only true for $z\simlt 1$. At higher redshifts, the fraction of FIR emitters of higher radio luminosities increases and by $z\simeq 2$ there is no dependence of such fraction on radio luminosity. 
\item Similarly, in agreement with the results of Magliocchetti et al. (2014), FIR emitters are found preferentially associated with galaxies of low, M$_*\simlt 10^{11}$ M$_\odot$, stellar masses only in the local universe. At higher redshifts such an effect loses its importance, and by $z\simeq 2$ galaxies of all stellar masses have the same chances of hosting a FIR-active, radio-selected AGN.
\item Also the distributions of the ages of the hosts of FIR-active and FIR-quiet AGN are indistinguishable from each other at redshifts $z\simgt 2$. More locally, these two classes of sources evolve differently and by $z\simlt 1$ FIR-quiet sources are found preferentially associated with older galaxies. 
\item As it was in Magliocchetti et al. (2014) and (2016), we find also in this case that FIR emission is entirely to be attributed to intense episodes of star-formation ongoing within the host galaxy. In this work we have shown that these episodes are so intense that the FIR luminosities of radio-selected AGN are on average higher than those of star-forming galaxies selected from the same radio sample in a consistent way. Furthermore, the distributions of stellar masses and star formation rates for these two classes of sources clearly show that FIR-active radio-selected AGN are on average not only more FIR-bright, but also more massive than star-forming galaxies. However, once again, this is only true in the local, $z\simlt 1$, universe. At higher redshifts the hosts of FIR-bright radio-selected AGN and star-forming galaxies are indistinguishable from each other. 
\end{enumerate}

The picture which therefore emerges from our analysis is that of a substantial similitude amongst galaxies which host a radio-active AGN, independent of whether the AGN phenomenon is associated with concomitant star-formation within the host, and also of a similitude between the hosts of radio-selected AGN and those of radio-selected star-forming galaxies. However, this is only true at high, $z\simgt 1-1.5$, redshifts. In the more local universe these similitudes break down and likewise star-forming galaxies, also radio-selected AGN associated with FIR emission preferentially inhabit hosts which are smaller and younger than those characterising the whole radio-active AGN parent population. Furthermore, such galaxies are preferentially occupied by AGN of relatively low radio luminosities.

Lastly, we have investigated the properties of radio-selected AGN which also show signatures for AGN emission at other wavelengths. We find that about 26\% of the radio-AGN belonging to our sample also emit in the X-ray, while $\sim 20$\% are also active in the MIR band. Both these percentages rise to $\sim 30$\%  in the case of radio-selected AGN which are also associated with star-formation within the host galaxy.

Our results indicate that while the sub-class of X-ray emitting AGN does not exhibit any sensible difference with respect to the whole radio-selected AGN population, the same is not true for those radio-emitting AGN which also show signatures for AGN activity in the MIR waveband. In fact, we find that this latter class of sources preferentially inhabits galaxies which are on average younger, less massive and more active at FIR wavelengths than the parent radio-AGN population. 

We can therefore envisage an evolutionary track for radio-active AGN which associates MIR emission mainly with the early stages of the life-time of the radio source, when the host galaxy was young and not many stars were already in place. On the other hand, X-ray emission in radio-selected AGN seems to be mainly associated with a later stage of the evolution of these sources, whereby their hosts are relatively old and have already built most of their stellar mass. This is also true for radio-selected AGN which only show emission at radio wavelengths, as these two latter classes of sources seem to be indistinguishable from each other.

AGN selected at all wavelengths (radio, X-ray and MIR) should represent objects caught in a short-lived, transition phase, with likely still on-going star formation, and possibly associated with outflowing winds. Interestingly, in three out of four cases in which ionised outflows have been clearly detected through spatially resolved near infrared follow-ups of luminous X-ray or MIR-selected AGN in the COSMOS field (Perna et al. 2015a,b; Brusa et al. 2016), the sources are also revealed as radio-active AGN, and are indeed part of the sample presented in this work. 
This observational result therefore fits a scenario where radiative-driven winds, produced when the black hole is accreting at its maximum, induce shocks in the host galaxy, accelerating relativistic particles which can then emit in the radio band (e.g. Zubovas \& King 2012).


\section{Appendix}

\begin{figure}
\includegraphics[scale=0.35]{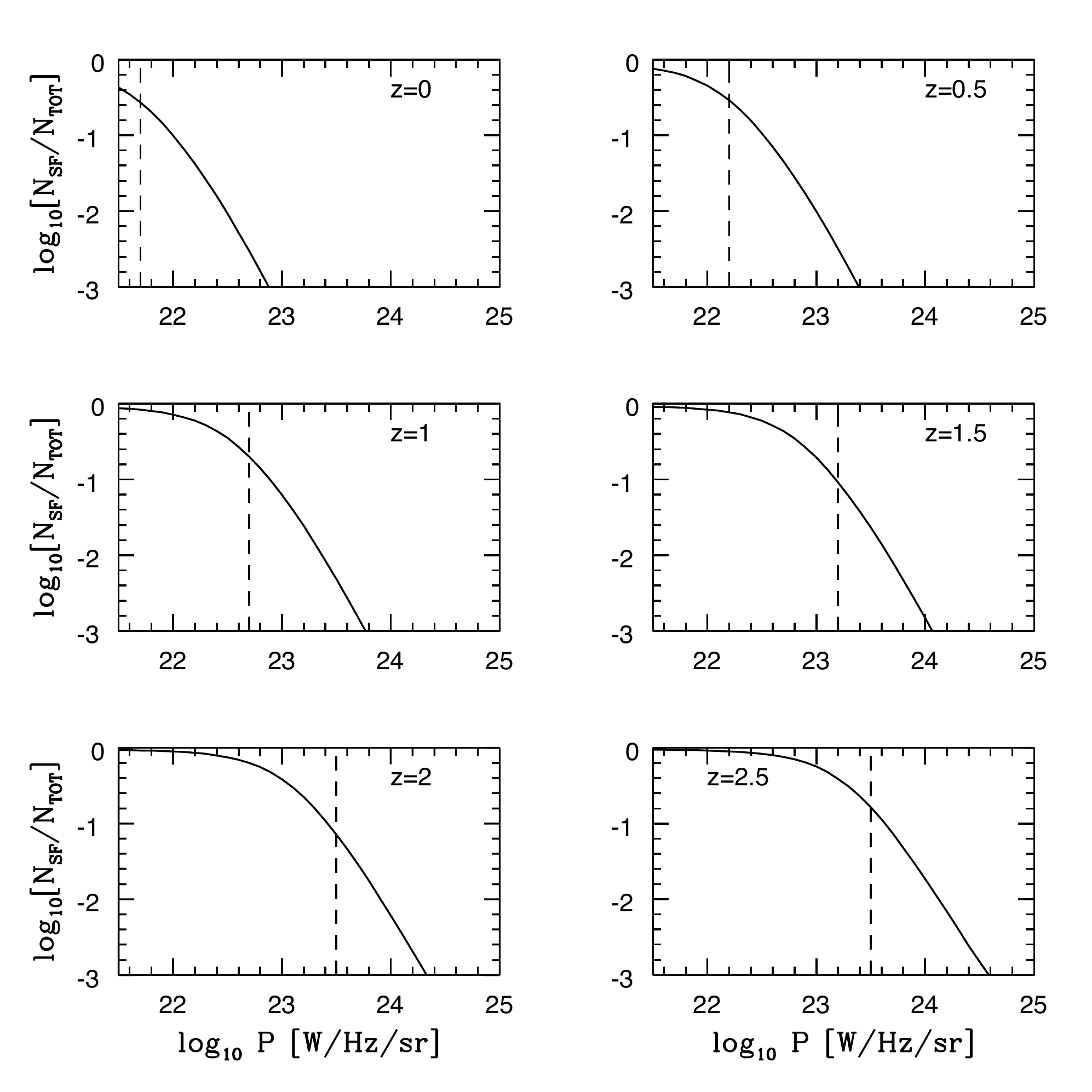}
\caption{Fraction of star-forming galaxies as a function of radio luminosity as derived from the luminosity function of McAlpine et al. (2013). The various panels represent different redshift intervals.
\label{fig:contamination}}
\end{figure}	

This Appendix is devoted to the discussion of the level of contamination of the AGN sample considered in the present work and will be divided into to parts: 
1) contamination due to the population of star-forming galaxies and 2) contamination due to AGN which owe their radio emission to star-forming activity rather 
than to the AGN itself.\\
\\
\noindent
1) Contamination from star-forming galaxies.\\

One possible source of contamination is that due to the un-removed presence of star-forming galaxies of radio luminosities higher than $P_{\rm cross}$ within the AGN sample. In order to assess the importance of such an effect, we have estimated the fraction of star-forming galaxies within our sample of radio-selected sources according to the radio luminosity function of McAlpine et al. (2013). Results at the various redshifts are presented in Figure \ref{fig:contamination}, where the vertical dashed lines represent the values of  $P_{\rm cross}(z)$ as defined in \S3.
As it is possible to notice, in all cases  star-forming galaxies tend to disappear very rapidly beyond $P_{\rm cross}$. In more detail,  in the local universe ($z=0$ and $z=0.5$ panels) the fraction of star-forming galaxies is $\sim 28$\% at $P_{\rm cross}$, while at $2 \cdot P_{\rm cross}$ this is already as low as $\sim 10$\%. At redshifts $z=1$ instead we find $N_{\rm SF}/N_{\rm TOT}\simeq 20$\% at $P_{\rm cross}$ and $\sim 6$\% at  $2 \cdot P_{\rm cross}$, while in the more distant, 
$z\ge 1.5$,  universe we have that the fraction of contaminants is about 10\% at $P_{\rm cross}$, and between 2\% and 4\% at  $2 \cdot P_{\rm cross}$.

\begin{figure*}
\includegraphics[scale=0.28]{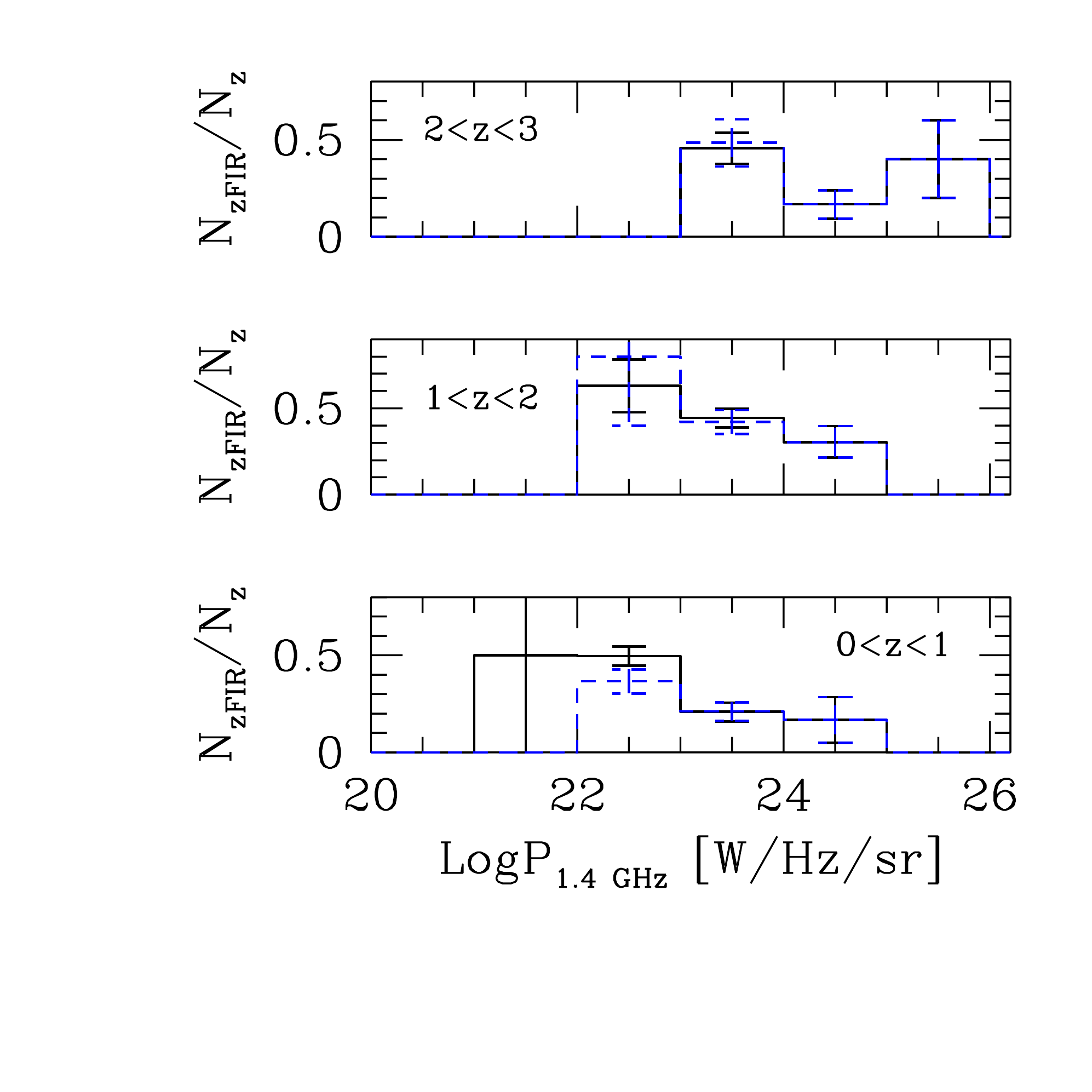}
\includegraphics[scale=0.28]{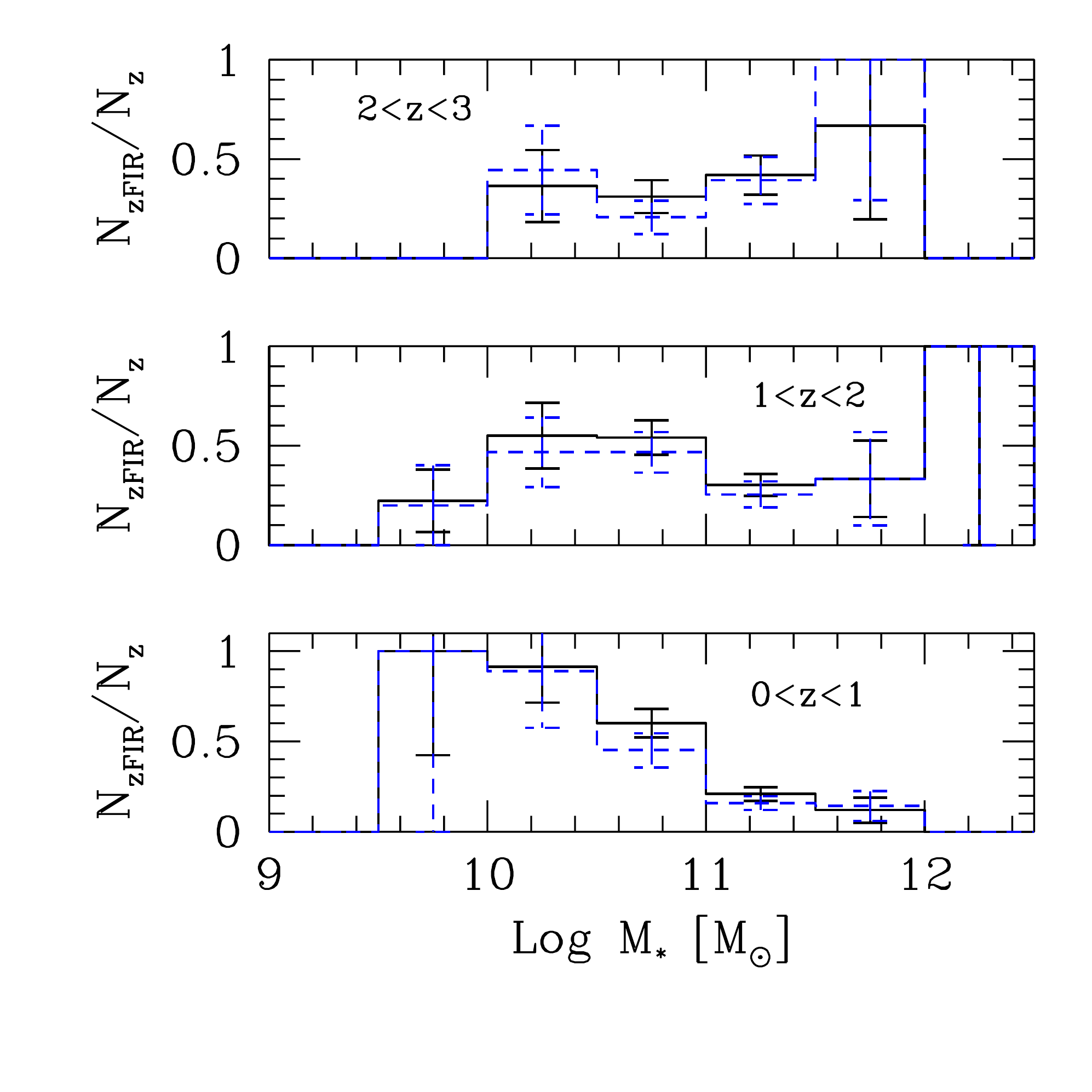}
\includegraphics[scale=0.28]{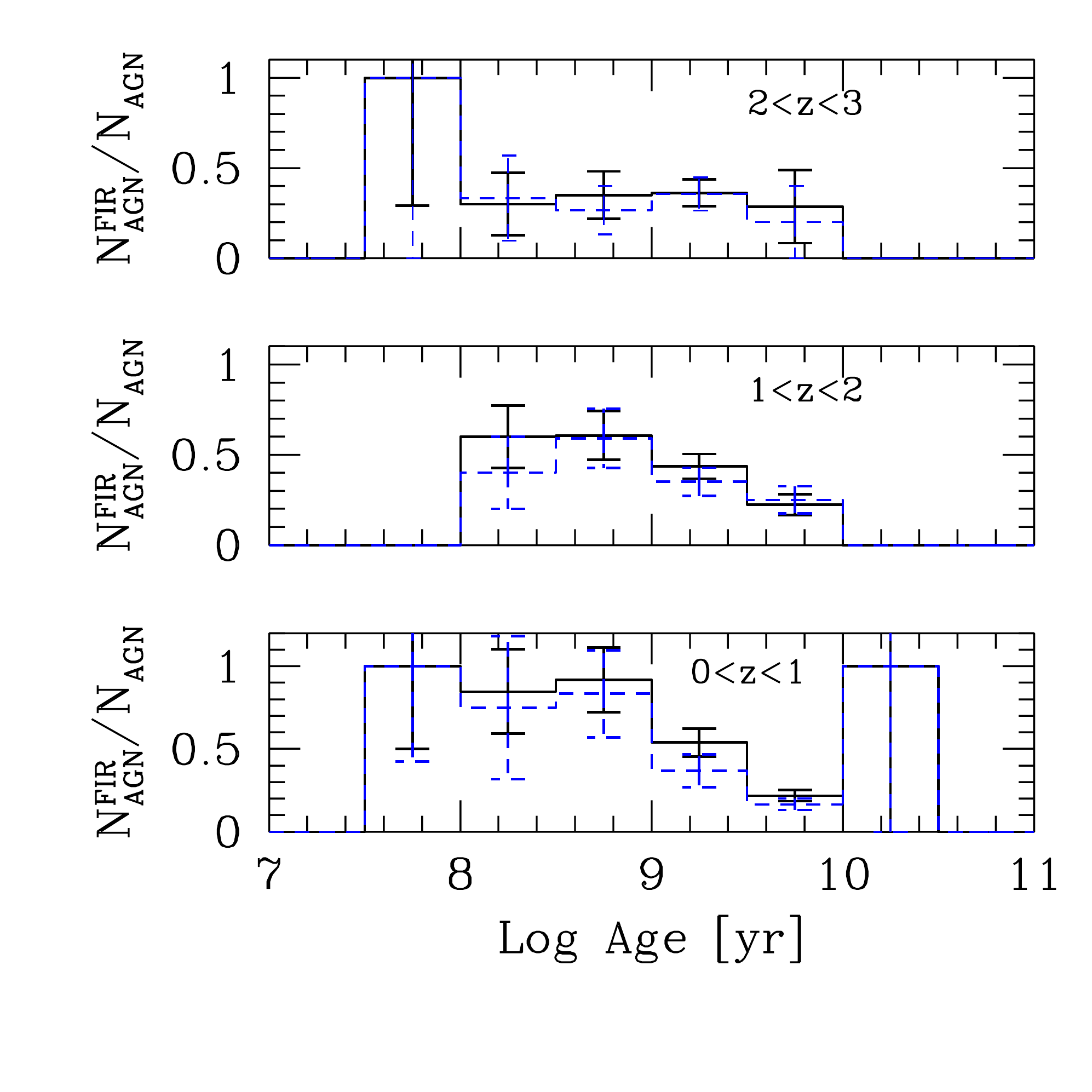}
\caption{Fraction of radio-selected AGN which also emit in the FIR bands as a function of 1) radio luminosity (left-hand panel), 2) stellar mass of the host (middle panel) and 3) age of the host (right-hand panel). The distributions are presented in three redshift ranges. In all cases, the solid histograms reproduce the results already presented in \S3, while the dashed ones show the variations obtained if one only considers sources brighter than twice the luminosity threshold $P_{\rm cross}$.
\label{fig:reresults}}
\end{figure*}

The above results clearly show that star-forming galaxies constitute a negligible fraction of the sample of radio-selected AGN already at radio luminosities which are twice as bright as the chosen threshold. In order to assess the robustness of the results presented throughout this paper, we have then performed once again our analysis by only including sources with  $P> 2 \cdot P_{\rm cross}$ (444 sources instead of 704, cfr \S3). As an example, the results for what concerns the distribution of AGN which are also active in the FIR bands as a function of radio luminosity (left-hand panel), stellar mass of the host (middle panel) and age of the host (right-hand panel) are presented in Figure \ref{fig:reresults}, whereby the solid histograms reproduce the results already presented in \S3 (Figures 2, 3 and 4), while the dashed ones what is obtained if we only concentrate on sources brighter than twice the luminosity threshold $P_{\rm cross}$. As it is clear from the Figure, there is no appreciable difference between the distributions of these two samples. In other words, the eventual presence of un-removed star-forming galaxies in the proximity of $P_{\rm cross}$ does not affect these or any other behavior or conclusion of the present work and also of the previous ones belonging to the same series (Magliocchetti et al. 2014; Magliocchetti et al. 2016).\\
\\
\noindent
2) Contamination from radio-quiet AGN.\\

\begin{figure}
\includegraphics[scale=0.4]{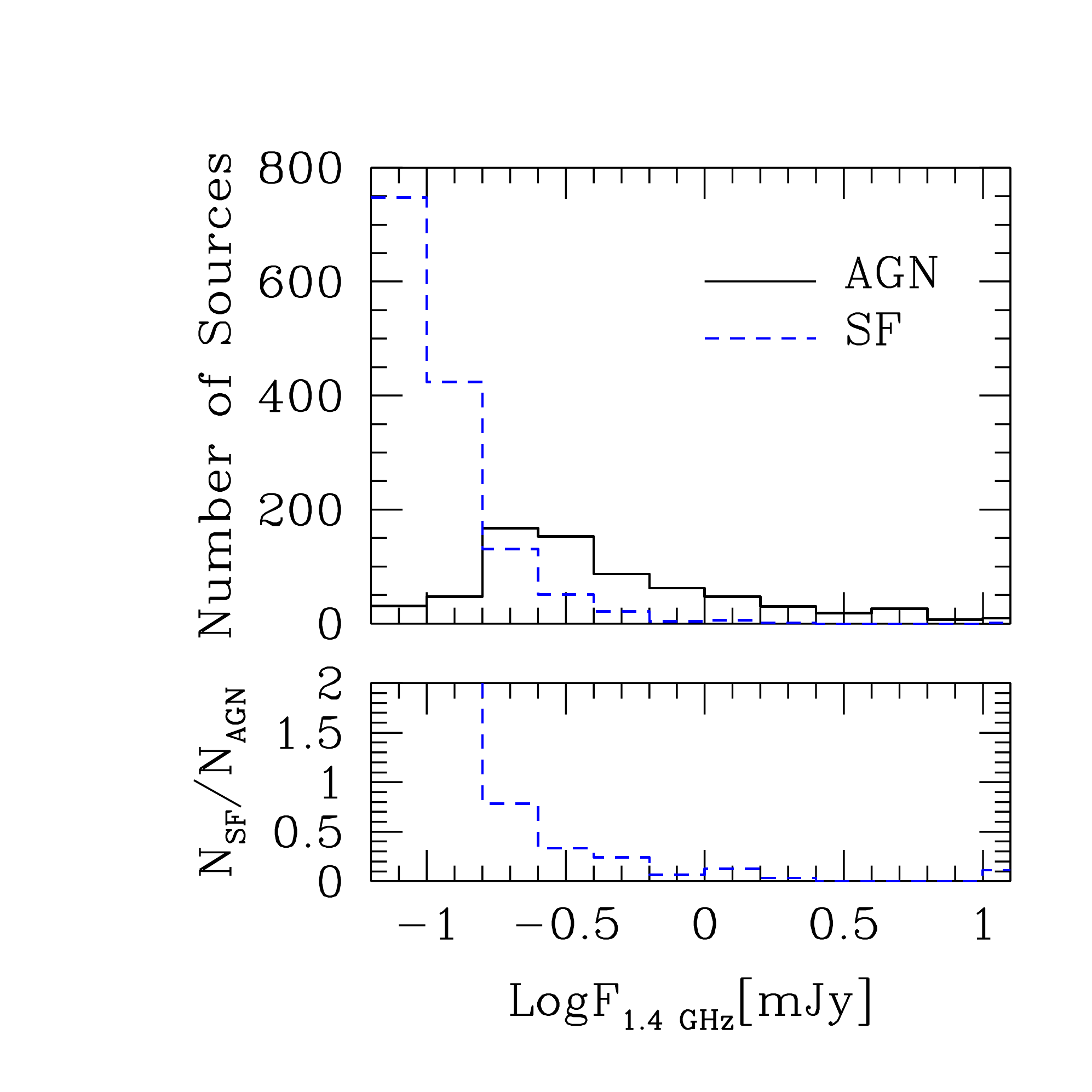}
\caption{Flux distribution of radio-emitting AGN (solid histogram) and radio-emitting star-forming galaxies (dashed histogram) selected by using the method highlighted in \S3. The bottom panel represents the ratio between these two quantities.
\label{fig:fractionAGNvsS}}
\end{figure}	

Another source of concern is represented by the possibility of contamination of our sample due to radio-quiet AGN which emit in the radio waveband only thanks to star-forming activity within the host galaxy. This issue was already discussed in Magliocchetti et al. (2014) who concluded that chances for such a contamination were extremely low. To further extend this point, we have computed the radio flux distribution for the two populations of radio-emitting AGN and star-forming galaxies as obtained by applying the method described in \S3. This is shown in Figure \ref{fig:fractionAGNvsS}, which clearly indicates that radio-selected AGN start  appearing in  the  COSMOS sample at fluxes $F_{\rm 1.4GHz} \simgt 0.15$ mJy and become the dominant population above $F_{\rm 1.4GHz} \simeq 0.5$ mJy.  
This  result  is  in  perfect  agreement  not  only  with  all  those found  in  the  literature, but  also  with  the  recent  ones  presented  by  Padovani  et al.  (2015),  which are based on the selection method of Bonzini et al. (2015) and with those presented in Smolcic et al. (2017).  Indeed, in agreement with our findings,   
Figure  1  of Padovani et al. (2015, but also cfr Figures 12 and 13 of Smolcic et al. 2017) clearly  shows  that  radio-emitting  AGN  (or  what  Padovani et al. 2015  call  "radio-loud  AGN" and Smolcic et al. 2017 call "radio-excess" sources; names vary  but  the  concept  is  the  same:  radio-detected  AGN  which  owe  their  emission  to accretion  processes  rather  than  to star-formation  within  the  host  galaxy)  beyond $F_{\rm 1.4GHz} \simeq 0.5$ mJy are  between  10  and  20  times  more  numerous  than  the  other classes of sources. Figure  1  of  Padovani  et  al.  (2015)   also  shows  (in  agreement  with  e.g.  the  results  of White  et  al.  2015)  
that  beyond  $F_{\rm 1.4GHz} \simeq 0.15$ mJy  the  contribution  to  the  total number  counts  of  the  population  of  radio-quiet  AGN  is  also  negligible.
 This  means that the chances that our AGN sample is contaminated by AGN which owe their emission to star-forming processes rather than accretion are negligible. \\
 In this respect, we would also like to point out the very recent results presented in Magliocchetti 
et  al. (2017),  which  show  that  the  clustering  properties  of  radio-selected  AGN  within 
the  COSMOS  area  (i.e.  basically  the  same sample  presented  in  this  paper)  are  in  full 
agreement with all those found in the literature (e.g. Magliocchetti et al. 2004; Brand et al. 2005; Wake et al. 2008; Fine et al. 2011; Linsday et al. 2014; Lindsay et al. 2014b; Retana-Montenegro \& Roettgering 2017). 
This would not be possible if some different   population  had   sensibly   contaminated   the   sample, 
since   the clustering  lengths  of  radio-quiet  AGN  (e.g. Porciani, Magliocchetti \& Norberg 2004; Retana-Montenegro \& Roettgering 2017) and  star-forming  galaxies  (e.g. Norberg et al. 2002; Magliocchetti \& Porciani 2003; Zehavi et al. 2005) are  much  smaller than  those  normally  measured  for  radio-loud  AGN  and  contamination  would  have 
brought the measured clustering signal below that which is typical of the radio-loud AGN population.\\
\\

\noindent
{\bf Acknowledgements}
MM and MB wish to thank  the DFG Cluster of Excellence 
'Origin and Structure of the Universe'
 (www.universe-cluster.de) for partial support during the completion of this work. 
 We also wish to thank the referee for his/her constructive comments which helped improving the paper.
 
 Based on data products from observations made with ESO Telescopes at the La
Silla Paranal Observatory under ESO programme ID 179.A-2005 and on data products produced by TERAPIX and the Cambridge Astronomy Survey Unit on behalf
of the UltraVISTA consortium.

¤¤

\end{document}